\documentclass[graphics,nofootinbib,aps,pra,howpacs,twocolumn,superscriptaddress]{revtex4-1}
\usepackage{graphicx}
\usepackage{dcolumn}
\usepackage{bm}
\usepackage{amsmath}
\usepackage{verbatim}
\usepackage{color}
\usepackage{subfigure}
\usepackage{times}
\usepackage{amsfonts,amssymb,graphics,epsfig}
\usepackage{tikz}
\usepackage{pgfplots}
\usepackage{physics}
\usepackage[colorlinks=false]{hyperref}

\renewcommand{\d}{\operatorname{d}}
\newcommand{\e}{\mathrm{e}}
\newcommand{\id}{\mathrm{I}}
\renewcommand{\i}{\mathrm{i}}

\newcommand{\Ch}{\mathrm{Ch}}
\newcommand{\avg}[1]{\left\langle{#1}\right\rangle}
\renewcommand{\Vec}[1]{\mathbf{#1}}
\newcommand{\figref}[1]{Fig. \ref{#1}}

\begin{document}
	\title{Topological numbers of Happer model with ``puzzling'' degeneracy in periodic magnetic field}
	
	\author{Yingkai Liu}
	\affiliation{School of Physics, Nankai University, Tianjin 300071, China}
	
	\author{Yifei Liu}
	\affiliation{School of Physics, Nankai University, Tianjin 300071, China}
	
	\author{Li-Wei Yu}
	\email{yuliwei@mail.nankai.edu.cn}
	\affiliation{Theoretical Physics Division, Chern Institute of Mathematics, Nankai University,
		Tianjin 300071, China }
	\begin{abstract}
		
 The Happer model, as the variation of Rabi-Breit model, describes the interactions between the total nuclear spin and the total electron spin-1 of the triplet dimer molecules of ${}^{87}\text{Rb}$. One interesting physical consequence of the Happer model is its puzzling degeneracy. In this paper, under the periodic driven magnetic field on total electron spin, the topological properties of the Happer model are present. Specifically, we calculate the Chern number of the system, both for the non-degenerate and degenerate cases.   We show that the Chern number is closely related to the total angular momentum of the system, instead of the electron spin.  Furthermore,  the perturbing spin-axis interaction term is also introduced for detecting the influence on the corresponding topological Chern number.   At last, in momentum space, we compare the Happer model with the topological semimetal in the sense of topological numbers.  In such model,  a ``magnetostatic shielding'' --like phenomena occurs.
 
	\end{abstract}
		\maketitle

	\section{Introduction}
	Degeneracy is an important concept in quantum mechanics.  Especially in condensed matter physics, degeneracy is closely  connected to the symmetry and quantum phase transition. Usually, the level crossing between ground state and the first excited state leads to the quantum phase transition in quantum many-body system\cite{zhu2006scaling,wolf2006quantum}. In resent years, the research of symmetry protected topological phase, such as topological insulators and topological superconductors\cite{moore2010birth,hasan2010colloquium,qi2011topological}, opens a new mind in connecting symmetries and topological phases as well as the degeneracy. For example, in topological insulators and semimetals\cite{volovik2003universe,armitage2018weyl}, the Dirac equation is properly used in describing the physical consequences, which generates degenerate Dirac point (or Weyl point)  connected with the magnetic monopole -- a topological construction associated to the Chern number.  The inner space is spinor space, which  is expressed by spin-$\frac{1}{2}$ operators. Recently, the topological consequences of higher spin-$k$ are also discussed\cite{bradlyn2016beyond,zhu2016triple,lv2017observation,hu2018topological}, whose corresponding degeneracy is $2k+1$-fold, rather than the 2-fold degenerate Weyl point.

	In this paper,  inspired by the level crossing in topological insulators and semimetals,  we employ the Happer model holding a puzzling degeneracy, and try to find some new physical consequences.
	
\begin{figure}
		\centering
		\subfigure[$S=1$, $L=\frac{1}{2}$:$2$-fold degeneracy\newline at $x=1$.]{
			\includegraphics[width=0.232\textwidth]{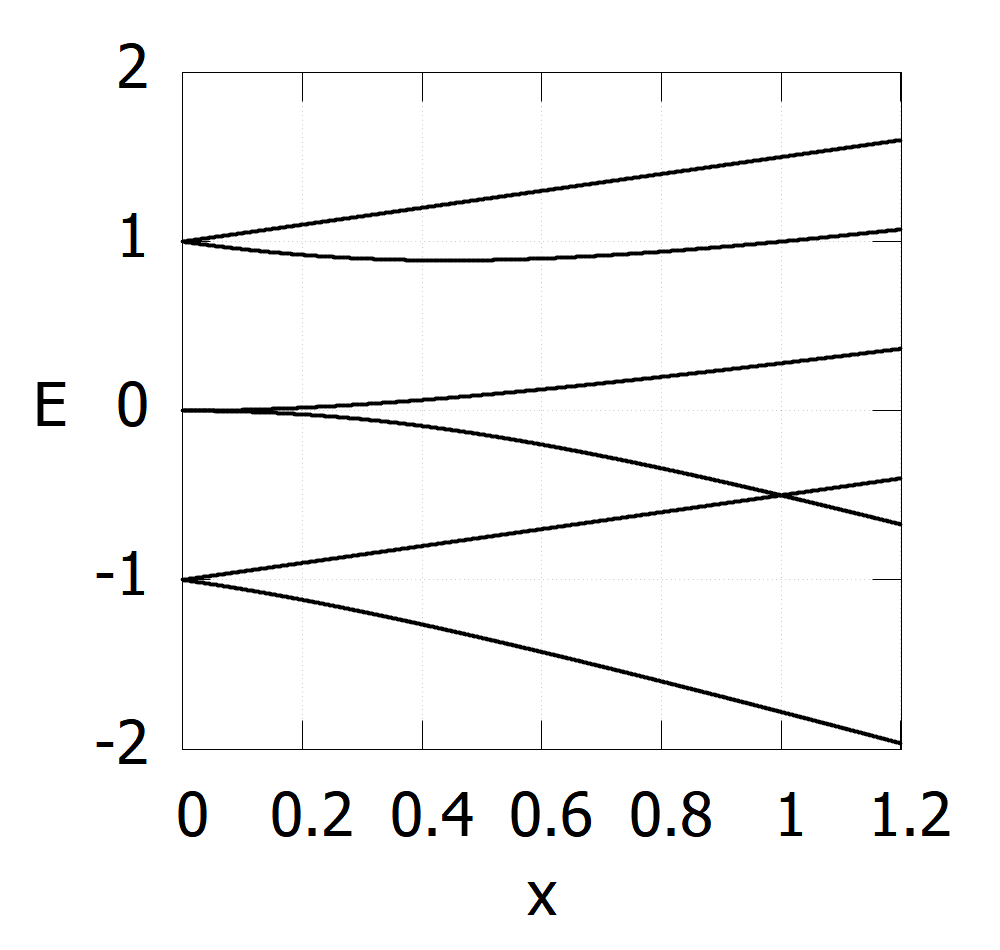}}
		\subfigure[$S=1$, $L=1$: $3$-fold degeneracy\newline at $x=\frac{2}{3}$.]{
			\includegraphics[width=0.23\textwidth]{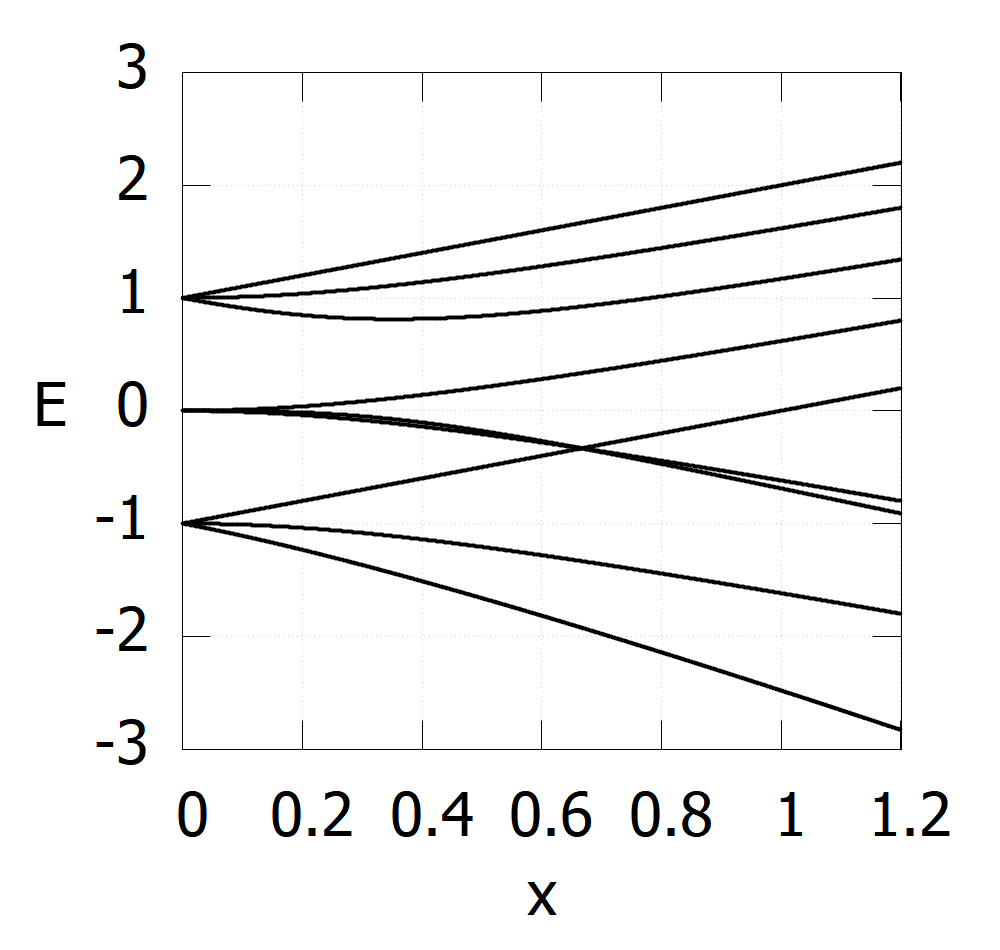}}
		\subfigure[$S=1$, $L=\frac{3}{2}$: $4$-fold degeneracy\newline at $x=\frac{1}{2}$.]{
			\includegraphics[width=0.232\textwidth]{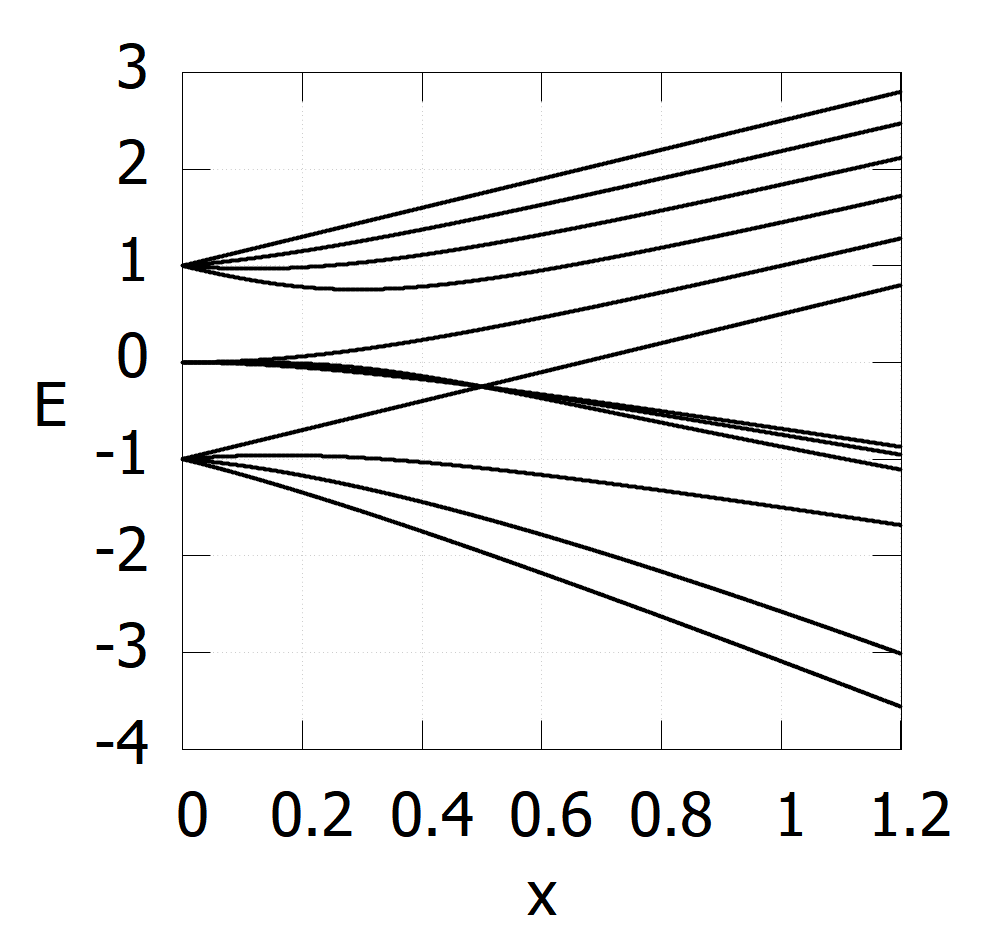}}
		\subfigure[$S=1$, $L=2$: $5$-fold degeneracy\newline at $x=\frac{2}{5}$.]{
			\includegraphics[width=0.23\textwidth]{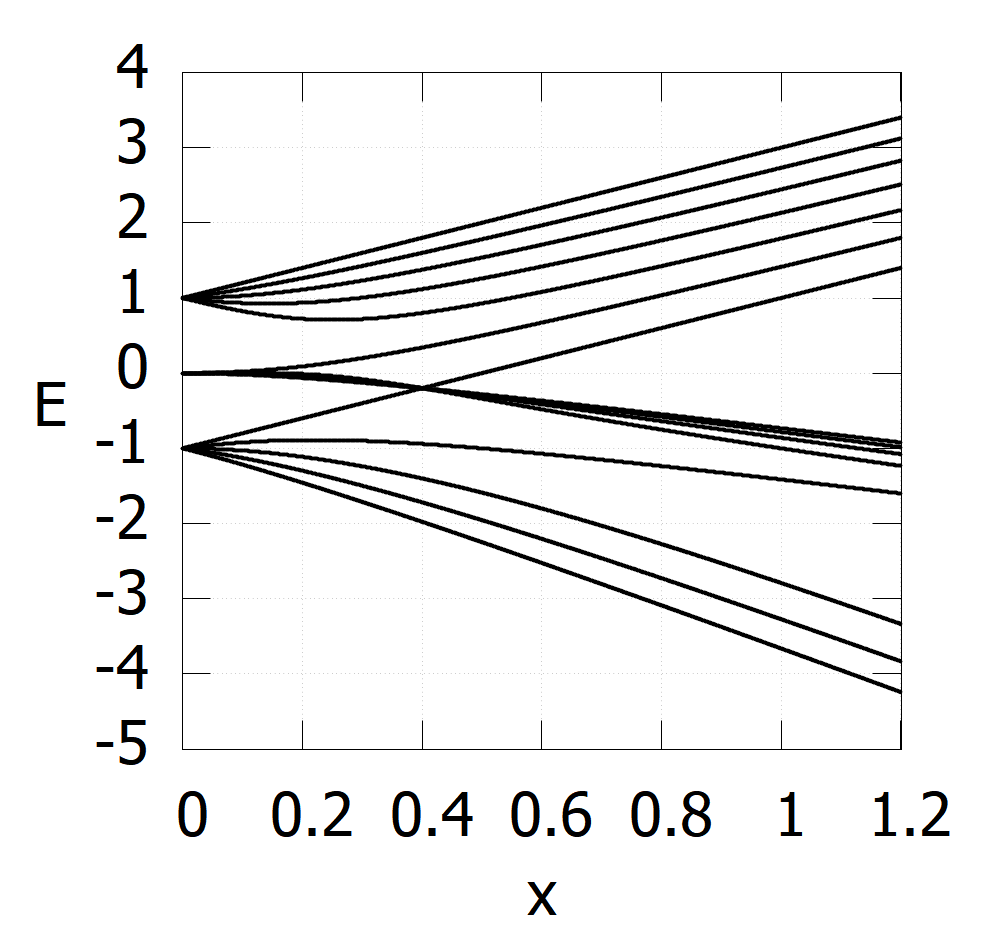}}
		\caption{Energy levels of the Happer model($y=0$) for different $L$. Later we will label the energy levels by $n$ from low to high at $x>\frac{2}{2l+1}$. \label{Fig:Ex}}
	\end{figure}

Let us first introduce the Happer model.  In explaining the experiment of spin-relaxation resonance of the alkali-metal Rb vapors, Happer \textit{et al.} \cite{erickson2000spin} proposed the following Hamiltonian
 \begin{equation}
	H= S_z + x\Vec{S} \cdot \Vec{L} + y \Vec{S} \cdot (3 \hat{\Vec{a}} \hat{\Vec{a}}-\id) \cdot \Vec{S},
	\label{Happer}
	\end{equation}
where $\Vec{S}$ is the spin-1 operator describing the triplet dimer of Rb$_2$ molecule,  $\Vec{L}$ is the nuclear spin-$L$, $\Vec{L}^2=L(L+1)$, and $\hat{\Vec{a}}$  is a unit vector along the direction of the internuclear axis. The first term on the RHS of Eq.~(\ref{Happer}) means the Zeeman splitting, the second term means the electro-nuclear spin coupling, and the third term means the spin-axis interactions originating from the collision of the Rb atoms with the coupling parameter $y$. This model can also be regarded as a variation of Breit-Rabi Hamiltonian \cite{breit1931measurement}. The interesting physical consequence occurs at $y=0$, where the Happer model holds a $2L+1$-fold degeneracy at $x=\frac{2}{2L+1}$. For example, the degeneracy is 3-fold for the nuclear spin $L=1$, while 5-fold for the nuclear spin $L=2$, see \figref{Fig:Ex}.

Since it was proposed, the puzzling $2L+1$-fold degeneracy has attracted much attentions. Because there seems no obvious symmetry in describing this degeneracy. The main focus is on detecting whether there is symmetry hidden in the degenerate point, and some progress has been made\cite{ge2002Curious,yuzbashyan2003extracting,gubser2005degenerate}. In \cite{ge2002Curious}, the authors proposed an approach of removing the degeneracy via the {\em Yangian} algebra, which can be regarded as an extension of the {\em Lie} algebras. In \cite{yuzbashyan2003extracting,gubser2005degenerate}, the authors systematically analyzed the degeneracy point and showed that it was protected by the $SU(2)$ group, but with the  non-standard representation -- the ``deformed'' raising operator  raised the lowest quantum number by 2 instead of 1, hence different from the normal raising operator.

As was mentioned above, the symmetry protected topological phases have attracted increasing attentions due to their exotic properties and potential applications. The two crucial properties of such topological non-trivial phases are the edge states and non-zero topological numbers, such as the winding number (for odd dimension) and Chern number (for even dimension). Especially the topological number reflects the energy band properties in momentum space, where the Hamiltonians are usually Dirac-like equations for spin-1/2 or  higher spin-$k$ cases.  The topological phase transition usually occurs in company with the closing of band gap, where the degeneracy appears.  

In this paper, motivated by the idea in connecting topological numbers with degeneracy as well as the phase transition, we explore the topological properties of  Happer model with $2L+1$-fold degeneracy. Specifically, we apply a periodic rotating magnetic field $\Vec{n}(\theta,\varphi)$ on the two-electron spin triplet $\Vec{S}=1$, and calculate the corresponding Chern numbers for both nuclear spins $L$=1 and $L$=2. Firstly,  we calculate the Chern number for non-degenerate levels by  defining the Berry curvature\cite{berry1984quantal} and integrating over the space.  Due to the total angular momentum $\Vec{S}+\Vec{L}$ conservation, the Chern number on the level is proportional to the total angular momentum quantum number $\Vec{n}\cdot(\Vec{S}+\Vec{L})$, instead of the electron spin operator $\Vec{n}\cdot \Vec{S}$. We then obtain the Chern number for ($2L$+1)-fold degenerate levels at $x=\frac{2}{2L+1}$ by defining the Wilczek-Zee curvature\cite{wilczek1984appearance} and making integration over the full space. ($\theta,\,\varphi$) The Chern number is shown equal to the sum of the Chern numbers of  the $2L$+1 crossing levels. From the properties of Happer degeneracy, it can be further deduced that the Chern number at the degenerate point is always 1 in spite of the $L$ values. After that, we add the spin-axis interaction term (see Eq.~(\ref{Happer})) and explore its influence on the degeneracy and topological properties of the Happer model. At last, the Happer model in projected subspace is compared with the topological semimetal model in the sense of topological numbers.

The paper is organized as follows. In Sec.~\ref{sec:II}, we show the topological properties of the Happer model without spin-axis interaction. In Sec.~\ref{sec:III},  we add spin-axis interaction in the system and detect its influence on the topological properties and degeneracy of the model. In Sec.~\ref{sec:IV}, we make comparisons between the projected Happer model with the topological semimetal model. In the last section, we make conclusions and discussions.

	\section{Topological number of Happer model without spin-axis interaction}\label{sec:II}
	
	By applying a rotating magnetic field on the spin $\Vec{S}$, the Happer model in Eq.~(\ref{Happer}) turns into the following form
	\begin{equation}
	H=\hat{\Vec{n}}_{B}\cdot \Vec{S} + x\Vec{S} \cdot \Vec{L} + y \Vec{S} \cdot (3 \hat{\Vec{a}} \hat{\Vec{a}}-\id) \cdot \Vec{S},
	\label{Eq:Full-Hamiltonian}
	\end{equation}
	where 
	\begin{equation*}
	\Vec{n}_B(\theta,\varphi)=\begin{pmatrix}
	\sin\theta\cos\varphi,\,
	\sin\theta\sin\varphi,\,
	\cos\theta
	\end{pmatrix} 	\end{equation*}
	is the unit vector along the direction of magnetic field, and $\Vec{S}$ and $\Vec{L}$ are spin operators. 
	
	In this section, we suppose the spin-axis interaction term be $y=0$. Then the Hamiltonian turns out to be
	\begin{equation}
	H=\hat{\Vec{n}}_{B}\cdot \Vec{S} + x\Vec{S} \cdot \Vec{L}.
	\label{Eq:Non-SA-Hamiltonian}
	\end{equation}
This Hamiltonian also exhibits a curious $(2L+1)$-fold level-crossing for $S=1$  at $x=\frac{2}{2L+1}$, like the $\Vec{n}_B=(0,\,0,\,1)$ case in \figref{Fig:Ex}. 
When the system includes spin-interaction($y\neq0$), the level crossing is converted to anti-crossing.

	We see that the Hamiltonian contains three parameters, $x$, $\theta$ and $\varphi$. In this paper, the corresponding Berry curvature or Wilczek-Zee curvature are defined in $(\theta,\varphi)$-space. While the tunneling parameter  $x$ leads to the change of Chern numbers on lower energy levels,  exhibiting like an order parameter describing phase transitions around $x=\frac{2}{2L+1}$.
	
	Now we calculate the topological Chern number of Happer model, including  both the non-degenerate and degenerate cases.	

	To start with, noting that the Hamiltonian processes a total angular momentum symmetry $J_{\Vec{n}_B}$, defined by the component of the total spin $\Vec J=\Vec S+\Vec L$ along the magnetic field $ \Vec{n}_B$,
	\begin{equation}
	J_{\Vec{n}_B} = \Vec{n}_B\cdot(\Vec{S}+\Vec{L}).
	\end{equation}	
It is easy to check that $J_{\Vec{n}_B}$ and $H$ commute. Hence each energy level corresponds to one $(\Vec{J}^2, J_{\Vec{n}_B})$ quantum number. Later we will show that the conserved total angular momentum sector $J_{\Vec{n}_B}$ coincides well with the Chern number.
	
	\subsection{$L=1$}
	
	For the $L=1$ case, the eigenenergies of Eq.~(\ref{Eq:Non-SA-Hamiltonian}) exhibit a three-fold degeneracy at the parameter value $x=\frac{2}{2L+1}=\frac{2}{3}$. In Fig. \ref{oppo1}, the states are numbered according to their energy levels from low to high at $x\rightarrow \infty$ by $n$. When $L=1$, there are totally $9$ states. The eigenstates $1,2,6,7,8,9$ are non-degenerate states for all $x$, while the states $3,4,5$ degenerate at $x=\tfrac{2}{3}$.
	
	\subsubsection{Non-degenerate states}
\begin{figure}
		\centering
		\subfigure[$S=1$, $L=1$:{Berry phase.}]{
			\includegraphics[width=0.232\textwidth]{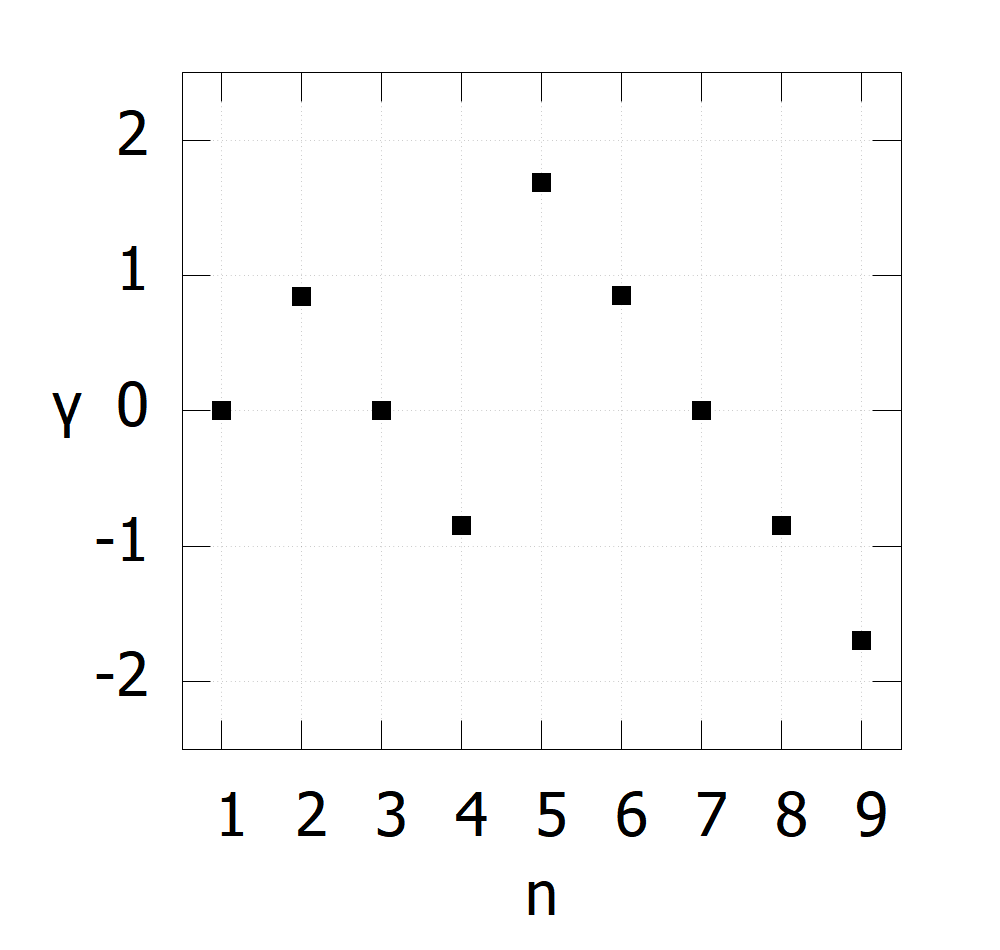}}
		\subfigure[$S=1$, $L=1$: Chern number.]{
			\includegraphics[width=0.232\textwidth]{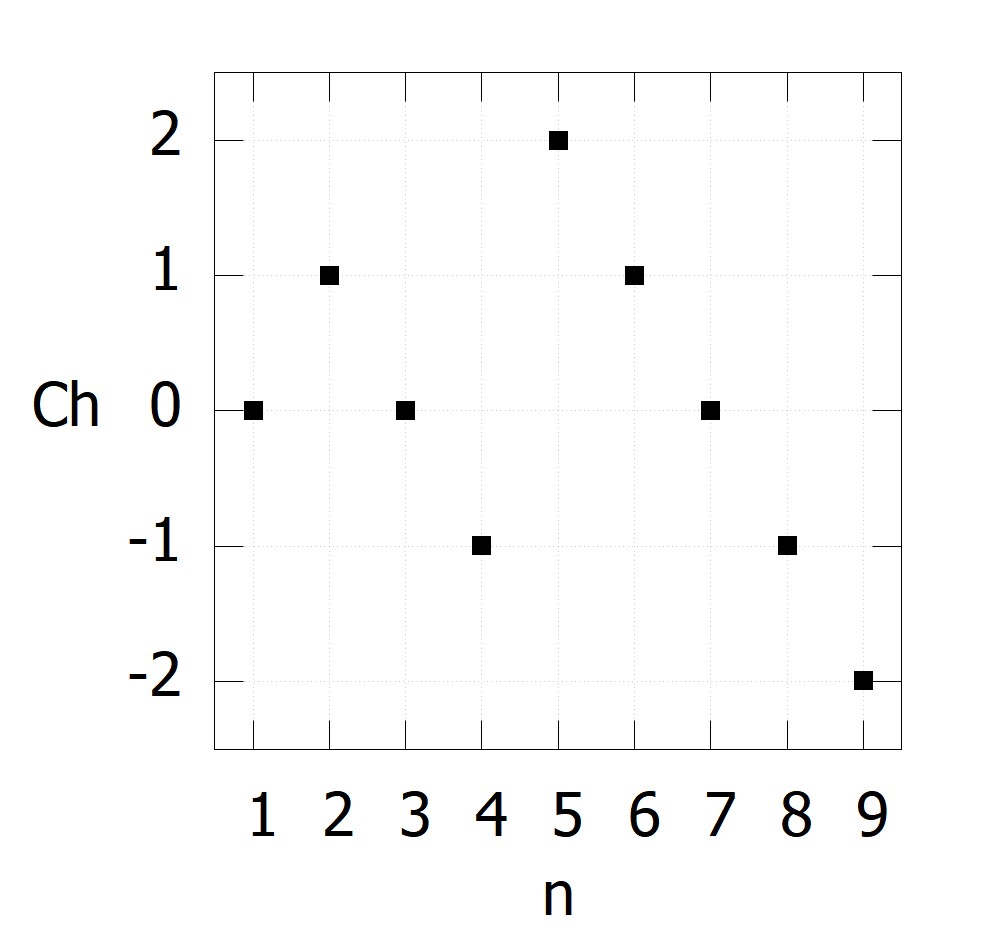}}
		\caption{Berry phase of the loop ($\theta=\frac{\pi}{6},\varphi=\omega t$) and Chern number at $x\ne\frac{2}{3}$.}
		\label{CG1}
	\end{figure}

\begin{figure}
		\centering
		\subfigure[$S=1$, $L=1$: $J_{\Vec{n}_B}$.]{
			\includegraphics[width=0.24\textwidth]{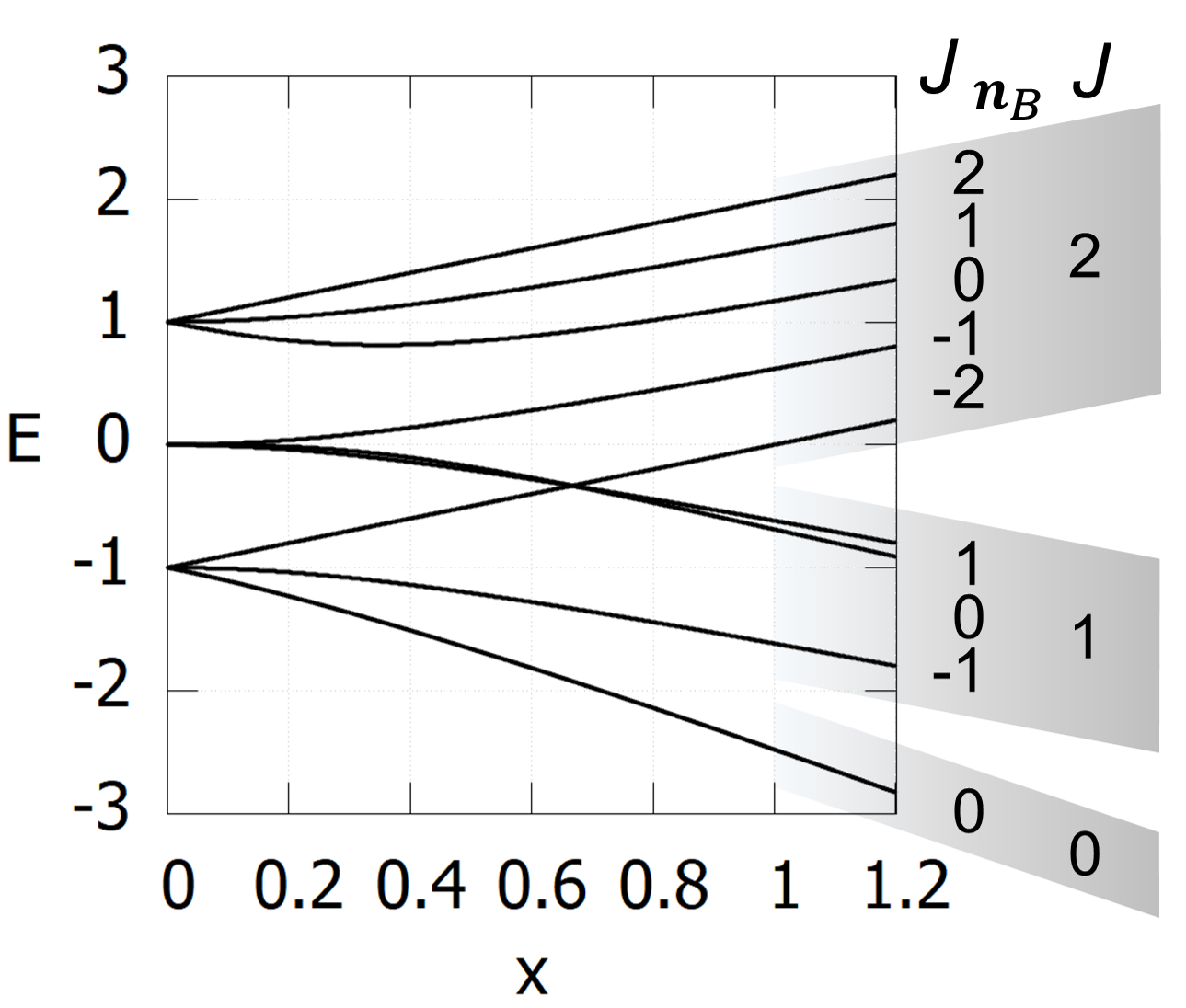}}
		\subfigure[$S=1$, $L=1$: Chern number and $J_{\Vec{n}_B}$.]{
			\includegraphics[width=0.226\textwidth]{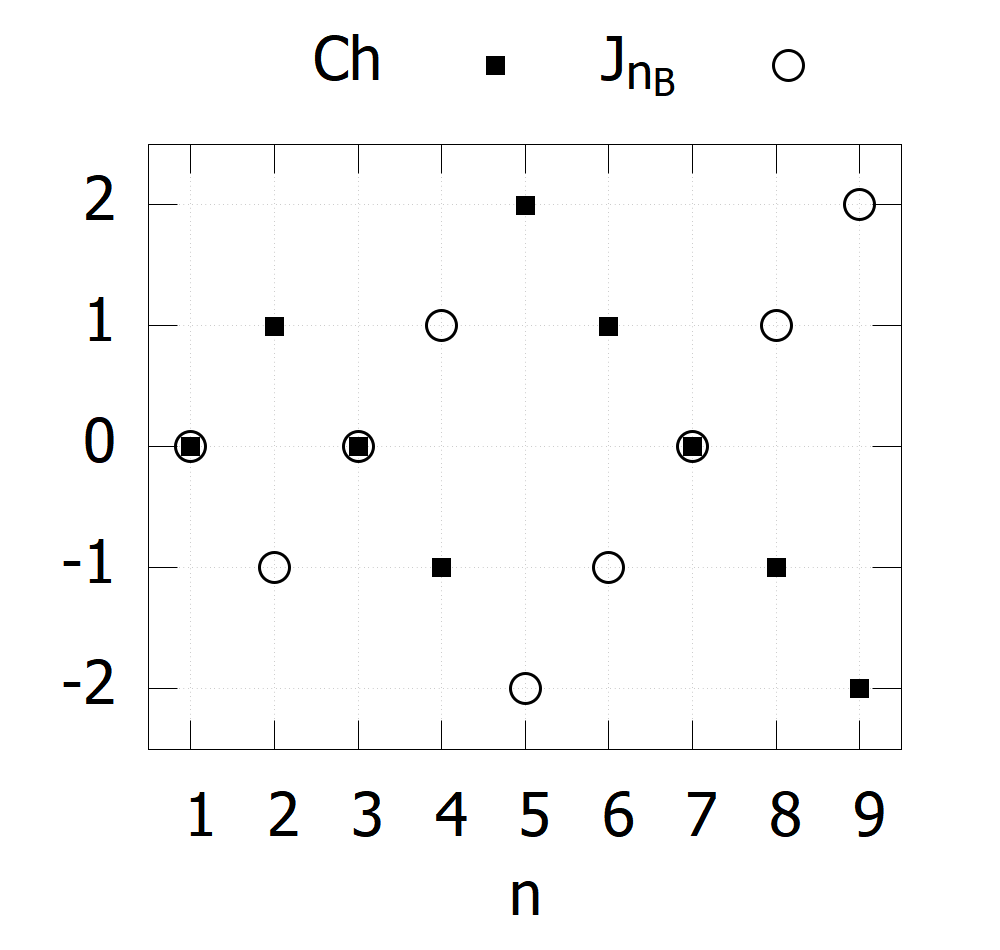}}
		\caption{Chern number is the opposite number of the $J_{\Vec{n}_B}$ of the state.}
		\label{oppo1}
	\end{figure}

	Here for completeness,  we first present the topological Chern numbers of the six non-degenerate levels. For each non-degenerate eigenstate $|\psi_i\rangle$ ($i$=1,2,6,7,8,9),  the Berry phase $\gamma^{(i)}$ and  Chern number $\Ch^{(i)}$ are defined as follows
\begin{eqnarray}
&& \gamma^{(i)}=\oint_{\ell}A^{(i)}_{\lambda}d\lambda,\\
&& \Ch^{(i)}=\frac{1}{4\pi}\int_{M}F^{(i)}_{\theta\varphi}\d \theta \wedge \d \varphi,
\end{eqnarray}
where 
\begin{equation}
\begin{split}
&F^{(i)}_{\theta\varphi}=(\partial_\theta A^{(i)}_\varphi-\partial_\varphi A^{(i)}_\theta),\\
&A^{(i)}_\lambda=i\langle\psi_i|\partial_\lambda|\psi_i\rangle, \quad (\lambda\in\{\theta,\varphi\})
\end{split}
\end{equation} represents the Berry curvature and connection of the evolving system under  the periodic magnetic field $\Vec{n}_B(\theta,\varphi)$. When these parameters change adiabatically, the non-degenerate states accumulate Berry phases along closed loops and Chern number over the 2D sphere $(\theta,\,\varphi)$. For instance, \figref{CG1} shows the Berry phases along the loop ($\theta=\frac{\pi}{6},\varphi: 0\rightarrow 2\pi$)  and the Chern numbers at $x\ne \frac{2}{3}$.
	Numerical calculations show that the Chern numbers are exactly the opposite number of $J_{\Vec{n}_B}^{(i)}$ of the states, see \figref{oppo1},
	\begin{equation}
	\Ch^{(i)}=-J_{\Vec{n}_B}^{(i)}=-\langle\psi_i|J_{\Vec{n}_B}|\psi_i\rangle.
	\end{equation}
Here recalling that we only apply the magnetic field on the subsystem electron triplet dimer $\Vec{S}$, however the physical result we obtain is similar to the case that the magnetic field applies on the total $\Vec{S}+\Vec{L}$ system.

	For more detailed insights, let us first show the Happer model with $x=y=0$ case, i.e. the precession of a spin $\Vec{S}$  under the adiabatic periodic magnetic field $\Vec{n}_B$,
	\begin{equation}
	H=\Vec{n}_B\cdot \Vec S.
	\end{equation}
Correspondingly, the Berry phase for the eigenstate $\ket{k;\theta,\varphi}$ is  
	\begin{equation}
	\gamma_k(C)=-k\Omega(C),
	\end{equation}
	where $\Omega(C)$ represents the solid angle accumulated from the adiabatic evolving in spin space,  and $k$ represents the eigenenergy (or the measurement of spin along the direction of magnetic field) of the state,
	\begin{equation}
	\Vec{n}_B\cdot \Vec S\ket{k;\theta,\varphi}=k\ket{k;\theta,\varphi}.
	\end{equation}
It is easy to check that the Chern number of the  eigenstate $\ket{k;\theta,\varphi}$ is 
	\begin{equation}
	\Ch_k=-k.
	\end{equation}

	\begin{table}[!ht]
		\centering
		\begin{tabular}[t]{lcc}
			\hline
			&  $\quad H=\Vec{n}_B\cdot \Vec S$  & $\quad H=x\Vec{S} \cdot \Vec{L} +\Vec{n}_B \cdot \Vec{S}$  \\
			\hline
			Conserved quantity&$\Vec{n}_B\cdot \Vec S$&$\Vec{n}_B \cdot( \Vec{S}+\Vec{L})$\\
			Eigenvalue&$k(=S_{\Vec{n}_B})$&$J_{\Vec{n}_B}$\\
			$\Ch$&$-k$&$-J_{\Vec{n}_B}$\\
			\hline
		\end{tabular}   
		\caption{Comparison between the two models.\label{czor}}
	\end{table}%
	
\begin{figure}[h]
		\centering
		\includegraphics[width=0.35\textwidth]{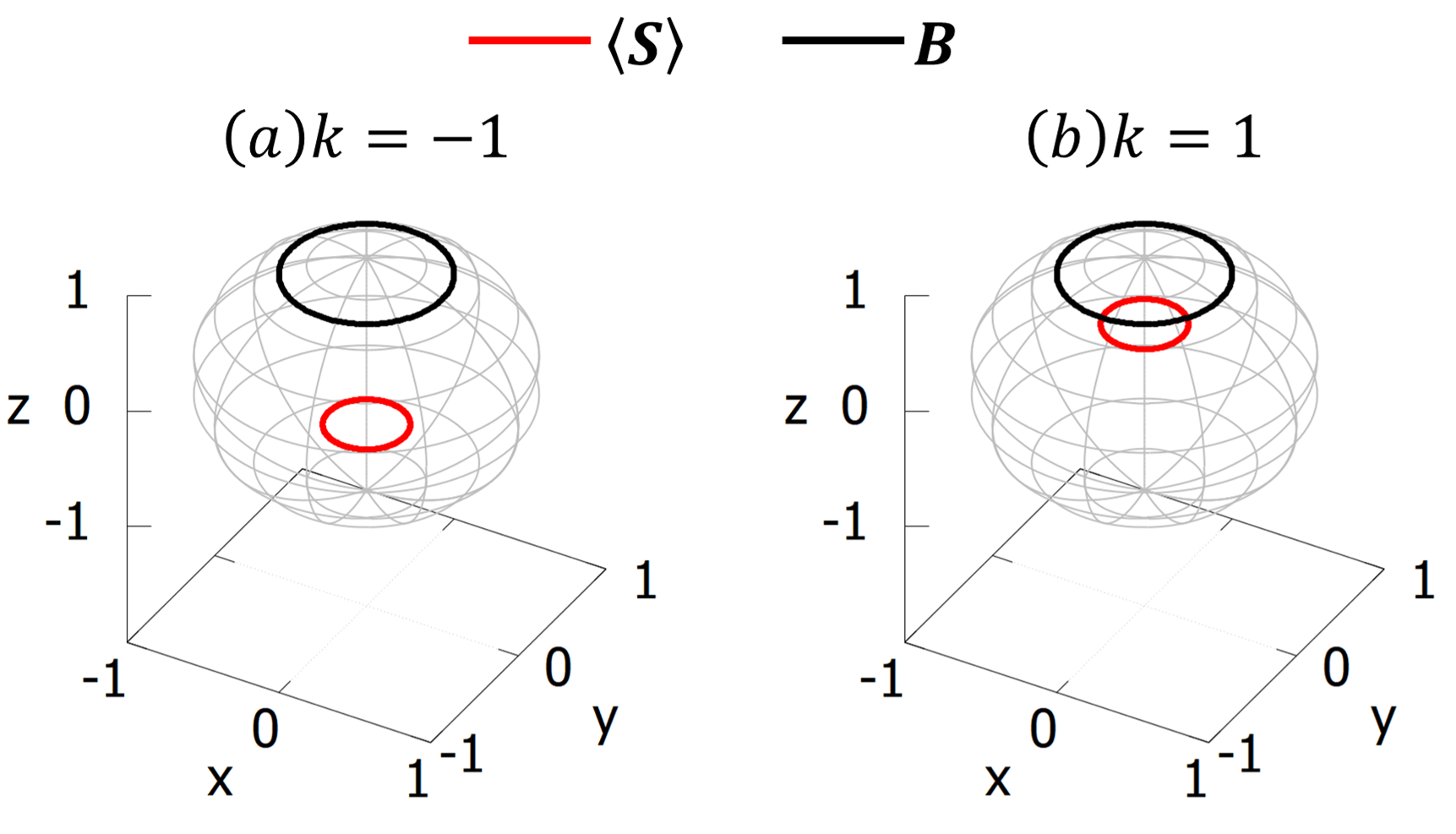}
		\caption{Motion of $\avg{\Vec{S}}$ for Hamiltonian $H=B\Vec{n}_B\cdot \Vec S$. (a) eigenstate $k=-1$; (b) eigenstate $k=1$.}
		\label{motion0}
	\end{figure}

\begin{figure}[h]
		\centering
		\includegraphics[width=0.35\textwidth]{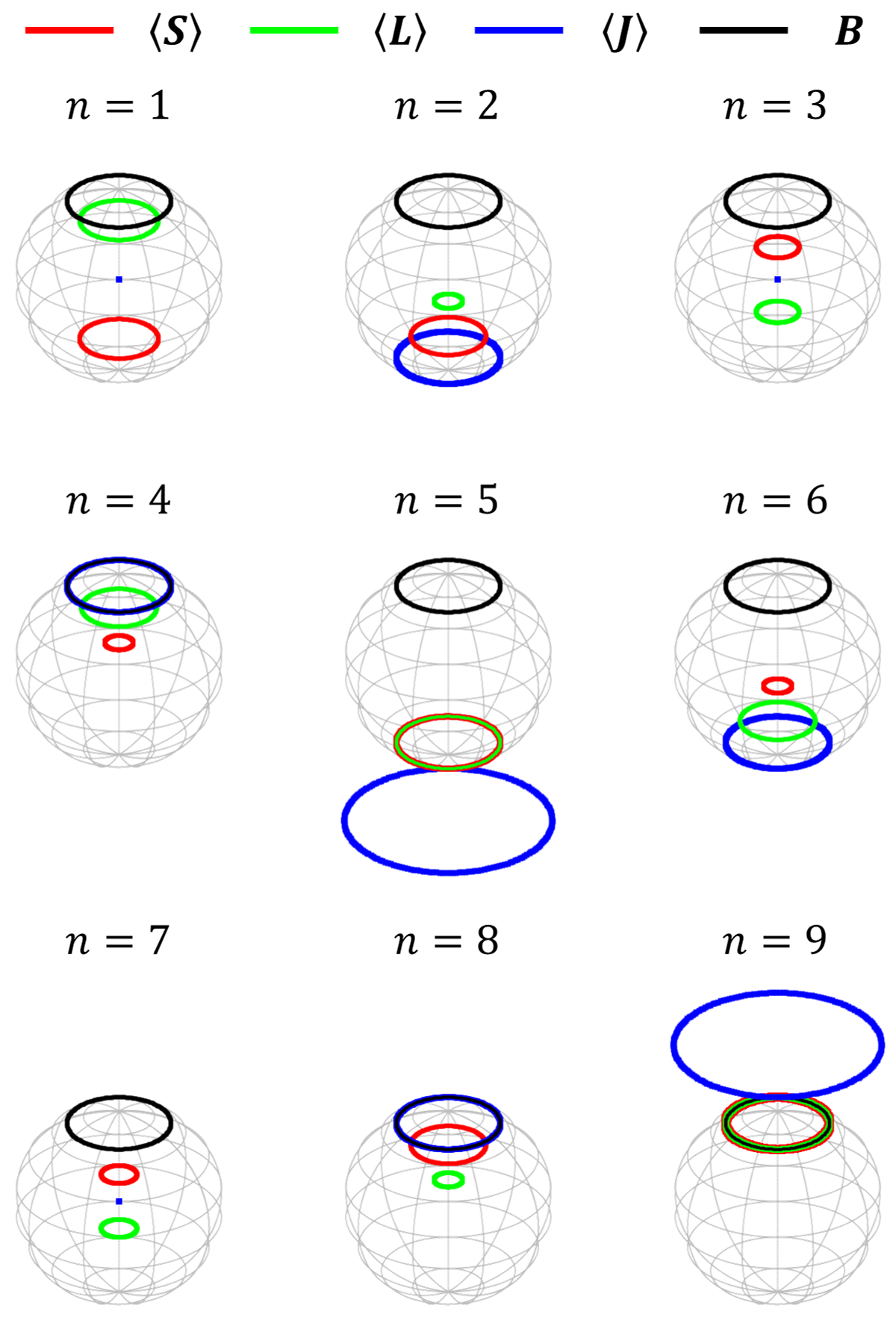}
		\caption{Motion of $\avg{\Vec{S}}$ and $\avg{\Vec{L}}$ for Hamiltonian $H=x\Vec{S} \cdot \Vec{L} +\Vec{n}_B \cdot \Vec{S}$.}
		\label{motion1}
	\end{figure}

{Table \ref{czor} shows the comparison between the two Hamiltonians. The Chern number of the systems governed by these Hamiltonians are precisely the conserved quantity, namely $S_{\Vec n _B}$ and $J_{\Vec n _B}$ respectively. This similarity is possibly due to the similarities of the kinetic behavior of the two systems - The adiabatic limit guarantees that the directions $\avg{\Vec{S}}$ and $\avg{\Vec{J}}$ always lie along or against the varying direction of the magnetic field,  see Fig. \ref{motion0} and Fig. \ref{motion1}.}


{In analogy to the case of $H=\Vec{n}_B\cdot \Vec S$, the absolute value of Berry phase is the oriented solid angle spanned by the corresponding trajectory of  $\avg{\Vec{J}}$.}

	
	\subsubsection{Degenerate states}
	The three  energy levels $n=3,4,5$ cross and  form a 3-fold degeneracy at the value $x=\frac{2}{3}$, $L=1$.  Here we deal with these three energy levels.  When $x\neq \tfrac{2}{3}$,  the three states are non-degenerate and only accumulate Berry phases, similar to the previous subsection. The interesting phenomena occurs at the degenerate point $x=2/3$. When the parameter $x$ approaches the degenerate point $x=\frac{2}{3}$ either from left or from right, the three states $n=3,4,5$ become degenerate  and the adiabatic  connection and curvature both become  $3\times 3$ matrices, which are described by Wilczek-Zee $SU(3)$ gauge field, a generalization of Berry $U(1)$ gauge field.  Generally, for an N-fold degenerate subspace, the matrix element $A^{mn}$ of  Wilczek-Zee connection $A(\theta,\varphi)$ is defined by
	\begin{align*}
	A ^{mn}_{\lambda} (\theta,\varphi) & = \i \bra{\psi_m,\theta,\varphi} {\partial_\lambda} \ket{\psi_n,\theta,\varphi},\quad (\lambda\in\{\theta,\varphi\}),
	\end{align*}
where the labels $m,n$ represent different  eigenstates in the degenerate space. The corresponding Wilczek-Zee curvature is expressed as
	\begin{align*}
	F_{\theta\varphi}^{mn}=\partial_\theta A^{mn}_{\varphi}-\partial_\varphi A^{mn}_{\theta}+i [A_{\theta},\, A_{\varphi}]^{mn}.
	\end{align*}
	Correspondingly, the Wilczek-Zee  geometrical phase $\gamma$ is the integral of the Wilczek-Zee potential along the closed loop $\ell=(\theta(t),\varphi(t))$,
		\begin{equation}
	\gamma  = \Tr\oint_\ell A_{\lambda}\d \lambda.
	\label{deggamma}
	\end{equation}
By integrating the Wilczek-Zee curvature over the full space, one obtains the Chern number formula of the system
	\begin{equation}
	\Ch  = \frac{1}{4\pi}\Tr\int_{S^2} F_{\theta\varphi} \d\theta \wedge \d \varphi.
	\label{degchern}
	\end{equation}
The trace $Tr$ ensures that the Chern number is $SU(3)$ gauge invariant.
	
	
	 After numerical calculations, we find that for degenerate states $n=3,4,5$ the Chern number $\Ch_{\text{deg}}$ at the degenerate point $x=\frac{2}{3}$ is exactly the sum of the Chern numbers of these states at non-degenerate $x$ values, i.e.
	\begin{equation}
	\Ch_{\text{deg}}\left(x=\frac{2}{3}\right) =\sum_{n=3}^5\Ch_n\left(x\ne\frac{2}{3}\right)=2+0-1=1.
	\end{equation}
	
	If we project the Hamiltonian in Eq.~\ref{Eq:Non-SA-Hamiltonian} into the subspace spanned by the three states $|\psi_n(x,\theta,\varphi)\rangle$ for energy level $n=3,4,5$, 
\begin{equation}\label{Hp}
H_P=P H P,\quad P=|\psi_3\rangle\langle\psi_3|+|\psi_4\rangle\langle\psi_4|+|\psi_5\rangle\langle\psi_5|,
\end{equation}	
then we can define the Chern number of the system $H'$ via the lowest energy band. When $x<2/3$, the Chern number of the lowest energy band is 2. While for $x>\frac{2}{3}$, the Chern number of the lowest energy band is 0. Hence, the level crossing point  $x=\frac{2}{3}$ can be regarded as the so called ``phase transition'' point in this sense. Detailed discussions about this will be present in Sec.~\ref{sec:IV}.

	\subsection{$L=2$}
	
	Similar to the $L=1$ case for three-fold degeneracy at $x=\frac{2}{3}$, in this subsection we discuss the topological properties of Happer model in $L=2$ case, where there are in total $15$ states in the system. In this case, there is five-fold degeneracy at the point $x=\frac{2}{5}$. The states are numbered according to their energy from low to high at $x\rightarrow \infty$ by $n$. The states labeled by $n=1,2,3,4,10,11,12,13,14,15$ are non-degenerate states, while the other states $n=5,6,7,8,9$ have degenerate energy level at $x=\tfrac{2}{5}$.
	
	\subsubsection{Non-degenerate states}

\begin{figure}[h!]
		\centering
		\subfigure[$S=1$, $L=2$: {Berry phase.}]{
			\includegraphics[width=0.232\textwidth]{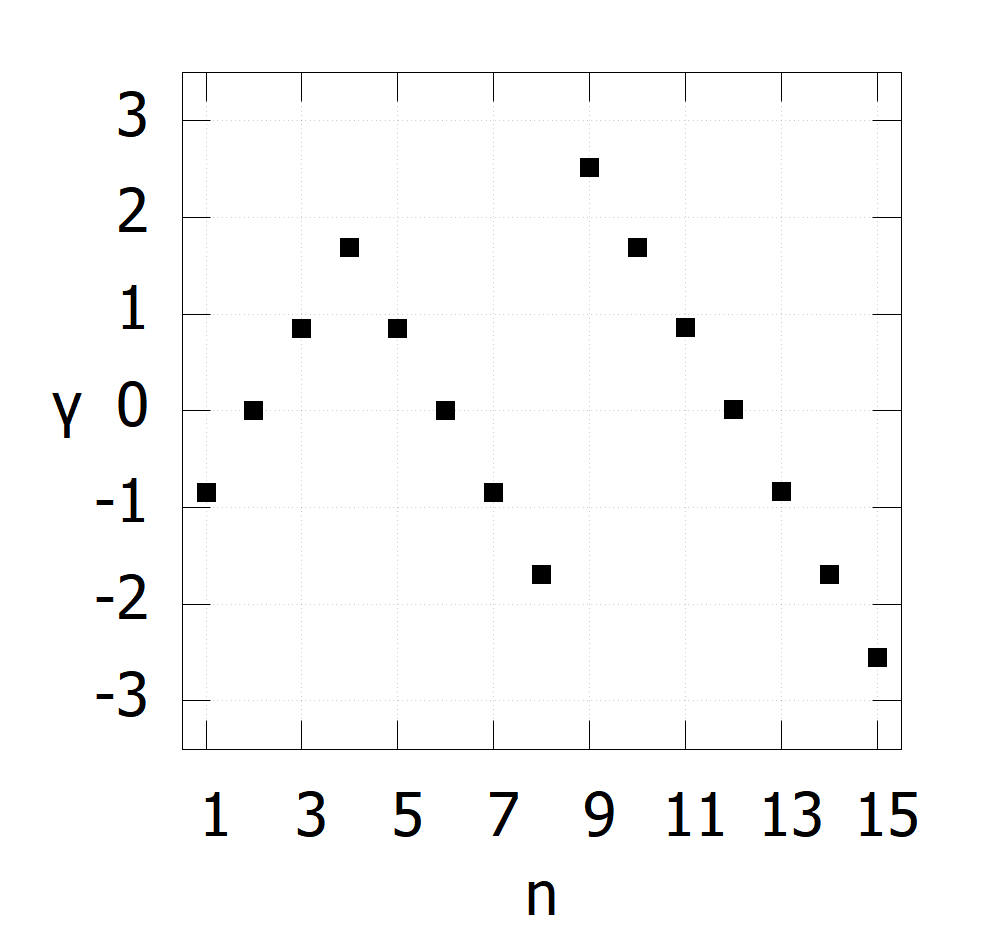}}
		\subfigure[$S=1$, $L=2$: Chern number.]{
			\includegraphics[width=0.232\textwidth]{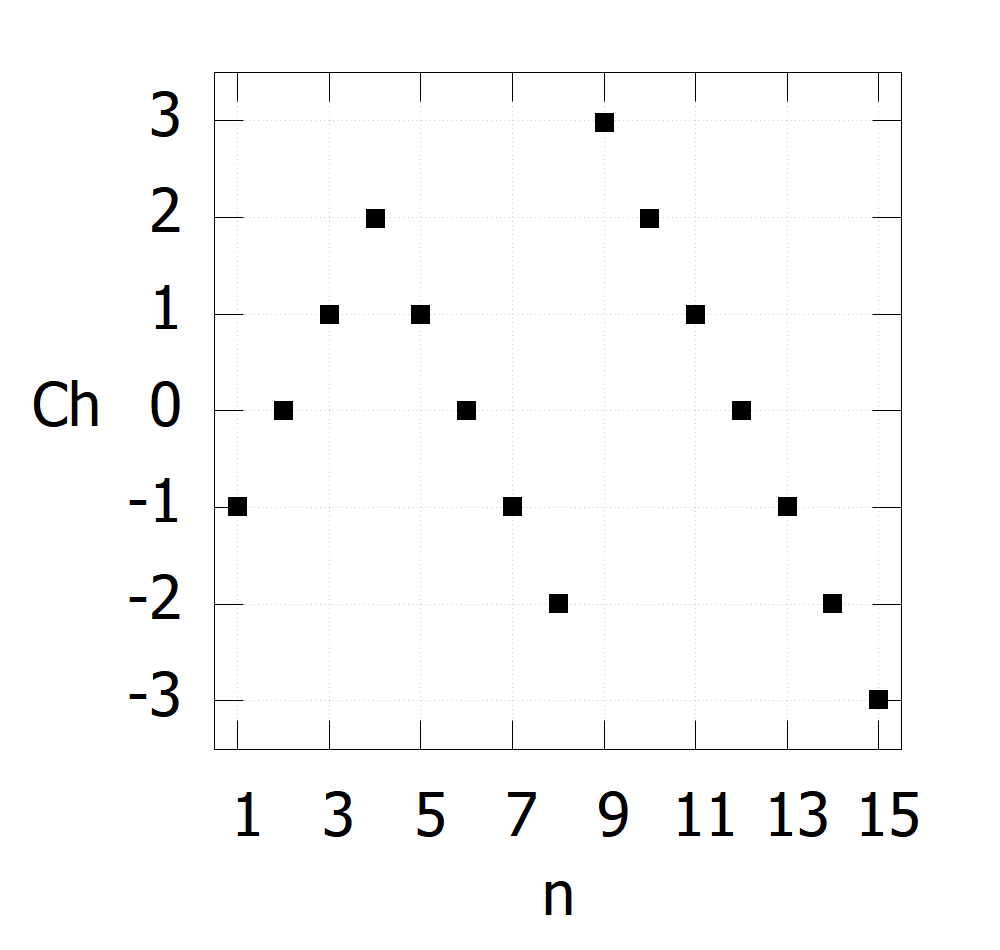}}
		\caption{Berry phase of the loop ($\theta=\frac{\pi}{6},\varphi=\omega t$) and Chern number at $x\ne \frac{2}{5}$.}
		\label{CG2}
	\end{figure}

\begin{figure}[h]
		\centering
		\subfigure[$S=1$, $L=2$: $J_{\Vec{n}_B}$.]{
			\includegraphics[width=0.24\textwidth]{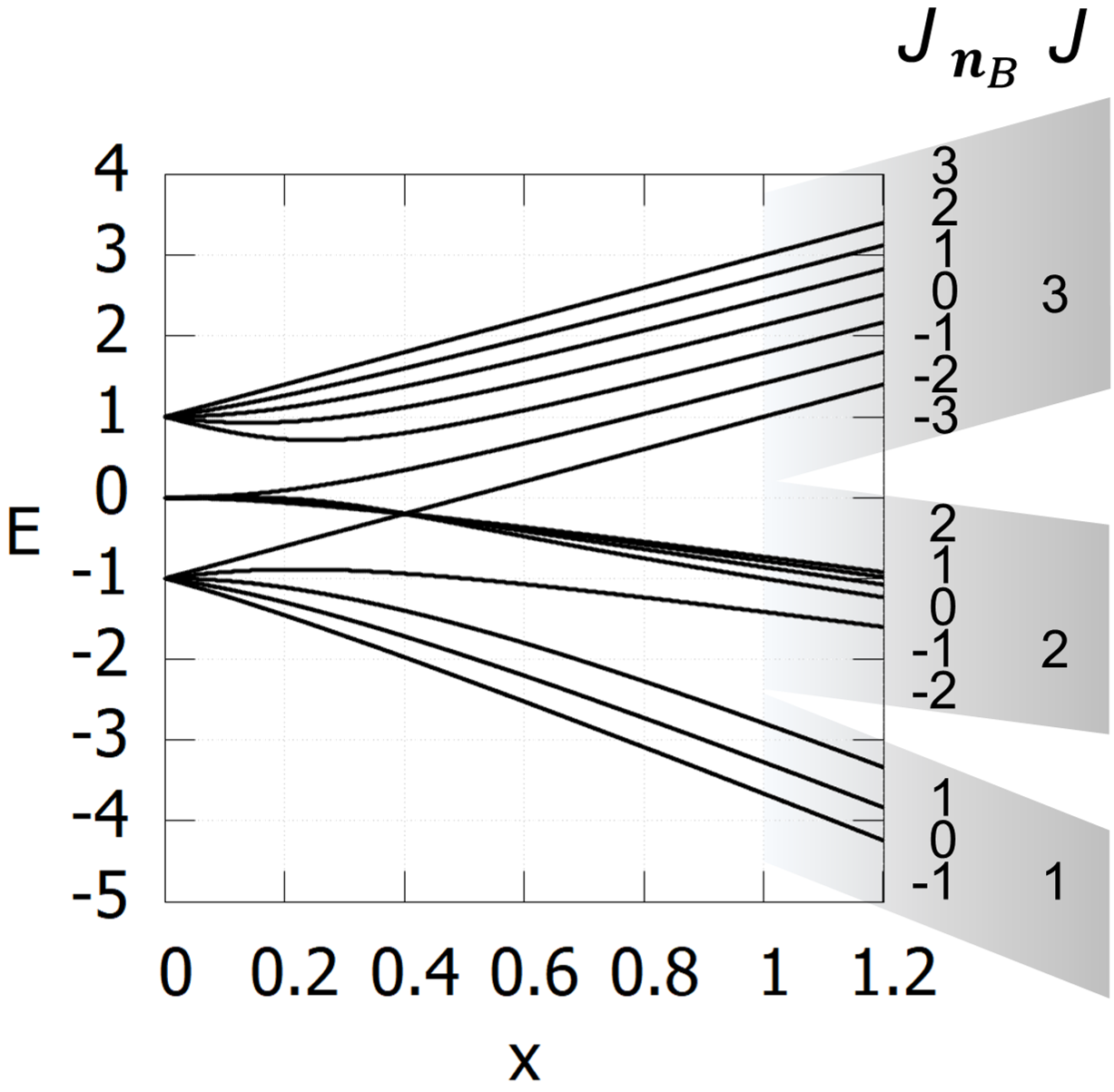}}
		\subfigure[$S=1$, $L=2$: Chern number and $J_{\Vec{n}_B}$.]{
			\includegraphics[width=0.226\textwidth]{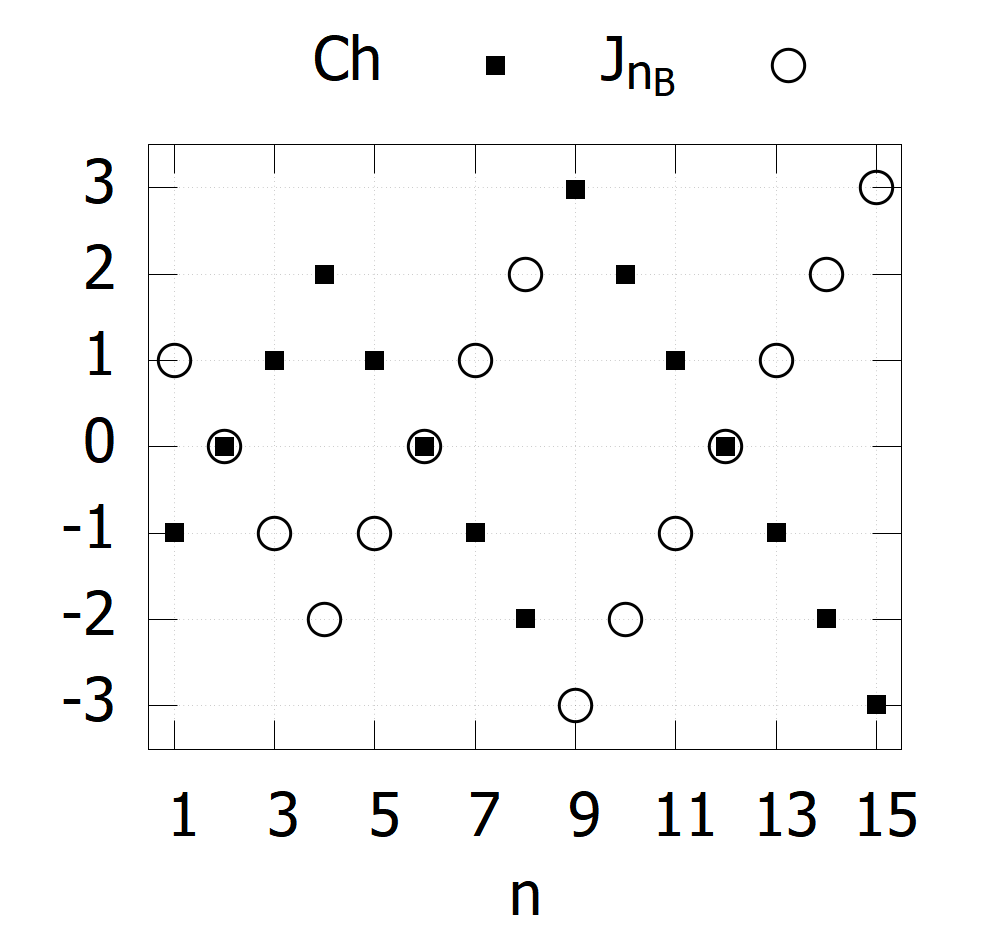}}
		\caption{Chern number is the opposite number of the $J_{\Vec{n}_B}$ of the state.}
		\label{oppo2}
	\end{figure}

\figref{oppo2}(a) shows the energy levels of our model for $L=2$. Here we focus on the ten energy levels without level crossing. The Hamiltonian is parametrized by the direction of the magnetic field, namely $\theta$ and $\phi$ in $\Vec{n}_B$. When these parameters change adiabatically, non-degenerate states will accumulate  Berry phases as well as Chern numbers.  For instance, the Berry phases at $x\ne \frac{2}{3}$ of the loop ($\theta=\frac{\pi}{6},\varphi=\omega t$) and the corresponding Chern numbers of the non-degenerate states are shown in Fig.\ref{CG2}.
	
	Numerical calculations show that the Chern numbers of these states are the opposite numbers of the $J_{\Vec{n}_B}$ of the corresponding states, see \figref{oppo2}, 
	\begin{equation}
	\Ch=-J_{\Vec{n}_B}.
	\end{equation}
	
\begin{figure}[h]
		\centering
		\includegraphics[width=0.48\textwidth]{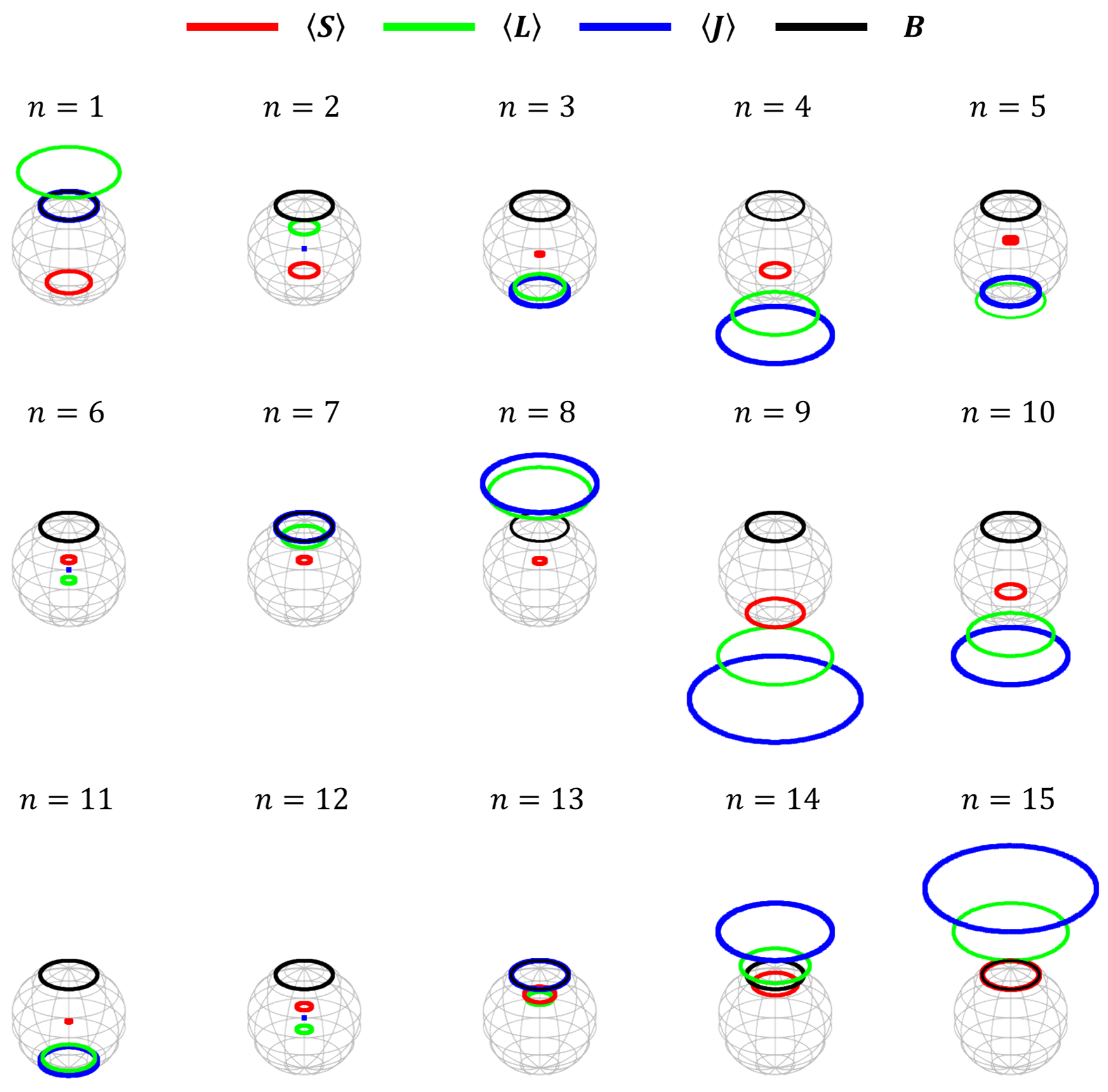}
		\caption{Motion of $\avg{\Vec{S}}$ and $\avg{\Vec{L}}$ for Hamiltonian $H=x\Vec{S} \cdot \Vec{L} +\Vec{n}_B \cdot \Vec{S}$.}
		\label{motion2}
	\end{figure}

	Similar to the $L=1$ case, the Chern numbers in $L=2$ system are also connected to the conserved quantity $J_{\Vec{n}_B}$. Indeed, The dynamics of $\Vec{J}$ in $L=2$ system is also similar to the Hamiltonian $H=\Vec{n}_B(t)\cdot \Vec{S}$, the concrete average values of $\Vec{S},\,\Vec{L}$ and $\Vec{J}$ are shown in Fig. \ref{motion2}. We can easily find that the absolute value of Berry phase is the {oriented} solid angle spanned by the trajectory of $\avg{\Vec{S}}$ or $\avg{\Vec{J}}$. 

	\subsubsection{Degenerate states}
	The states $n=5,6,7,8,9$ forms five-fold degeneracy at $x=\frac{2}{5}$. Here we discuss the topological behavior of the five states for both $x\neq \frac{2}{5}$ and $x=\frac{2}{5}$.  when the parameter $x\neq \tfrac{2}{5}$, each state holds $U(1)$ gauged Berry curvature  in periodic $(\theta,\varphi)$-space, and the Chern number is connected to the $J_{\Vec{n}_B}$ eigenvalue. When the parameter $x$ approaches the degenerate point $x=\frac{2}{5}$ either from the left or from right, the degenerate states form a 5d-space and hold $SU(5)$ gauged Wilczek-Zee curvature, 	
	
	\begin{equation}
	\gamma_{\text{deg}}\left(x=\frac{2}{5}\right) =\sum_{n=5}^9\gamma_n\left(x\ne\frac{2}{5}\right).
	\end{equation}
	where $\gamma_{\text{deg}}$ is the Wilzeck-Zee phase of the states of $n=5,6,7,8,9$ at degenerate point calculated by Eq.\ref{deggamma} and $\gamma_n$ is the Berry phase of the single state at non-degenerate point.
	
	By numerical calculations of integrating Wilczek-Zee curvature over the space $(\theta,\varphi)$, the Chern number $\Ch_{\text{deg}}=1$ at $x=\frac{2}{5}$. Similar to the $L=1$ case, it shows that for degenerate states $n=5,6,7,8,9$ the Chern number $\Ch_{\text{deg}}$ at the degenerate point $x=\frac{2}{5}$ is exactly the sum of the Chern numbers of these states at non-degenerate $x$ values, i.e.
	\begin{equation}
	\Ch_{\text{deg}}\left(x=\frac{2}{5}\right) =\sum_{n=5}^9\Ch_n\left(x\ne\frac{2}{5}\right)=3-2-1-0+1=1.
	\end{equation}
An interesting observation is that for $L=1$ and $L=2$ cases, both of the Chern numbers at the degenerate point $x=\frac{2}{2L+1}$ are $\Ch_{\text{deg}}=1$. Indeed, from the calculations of eigenstates for general $L$, we deduce that the Chern number at the degenerate point $x=\frac{2}{2L+1}$ is always $\Ch_{\text{deg}}=1$, i.e. independent of the concrete value of $L$.
	
	\section{Removing degeneracy under perturbations of the spin-axis interaction}\label{sec:III}
\begin{figure}[h]
		\centering
		\includegraphics[width=0.49\textwidth]{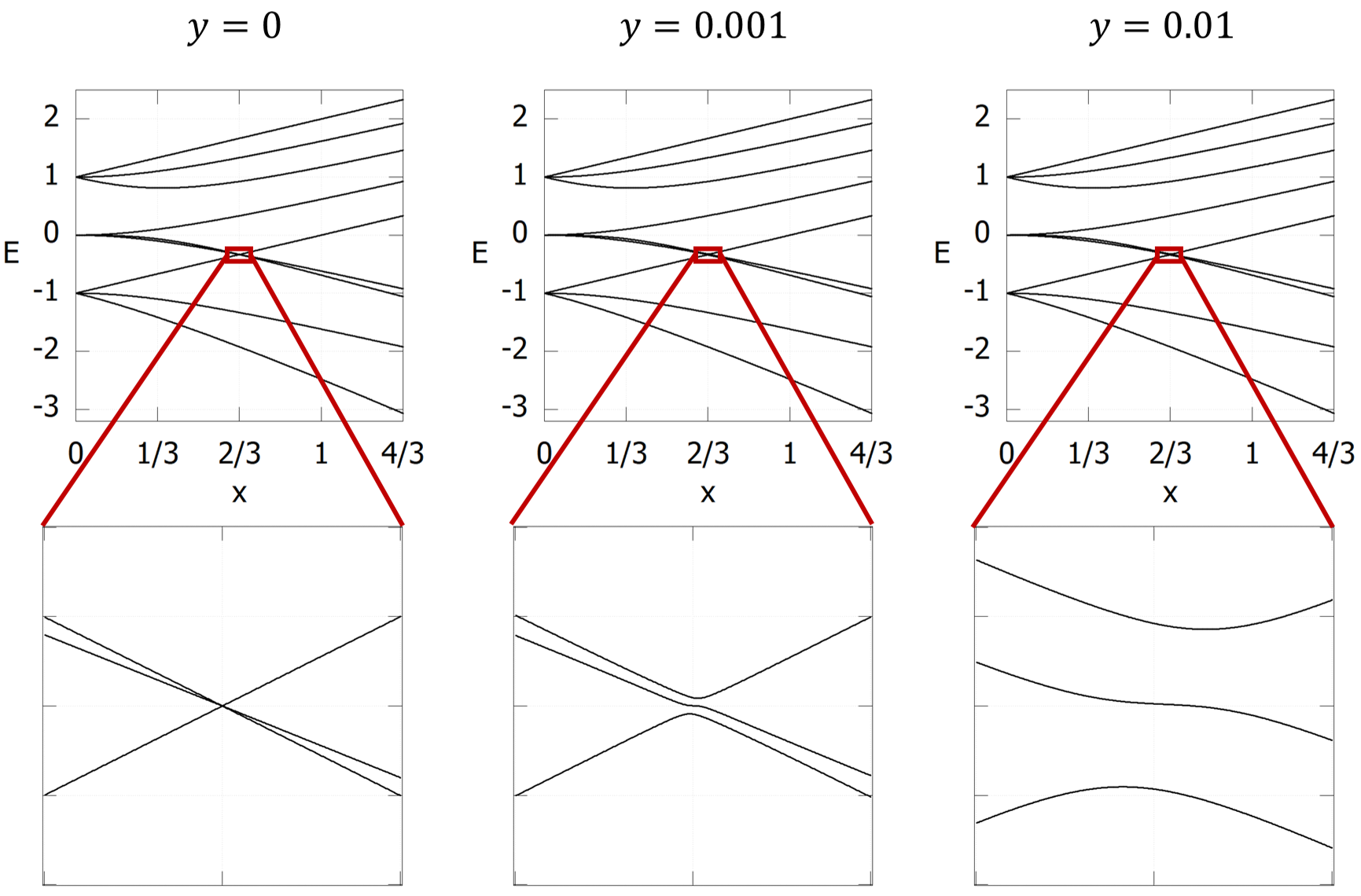}
		\caption{When y=0, the Hamiltonian has a $2L+1$ fold degeneracy. When $y\ne 0$, the crossing turns into the anti-crossing.}
		\label{spec1}
	\end{figure}

\begin{figure}[h!]
		\centering
		\includegraphics[width=0.49\textwidth]{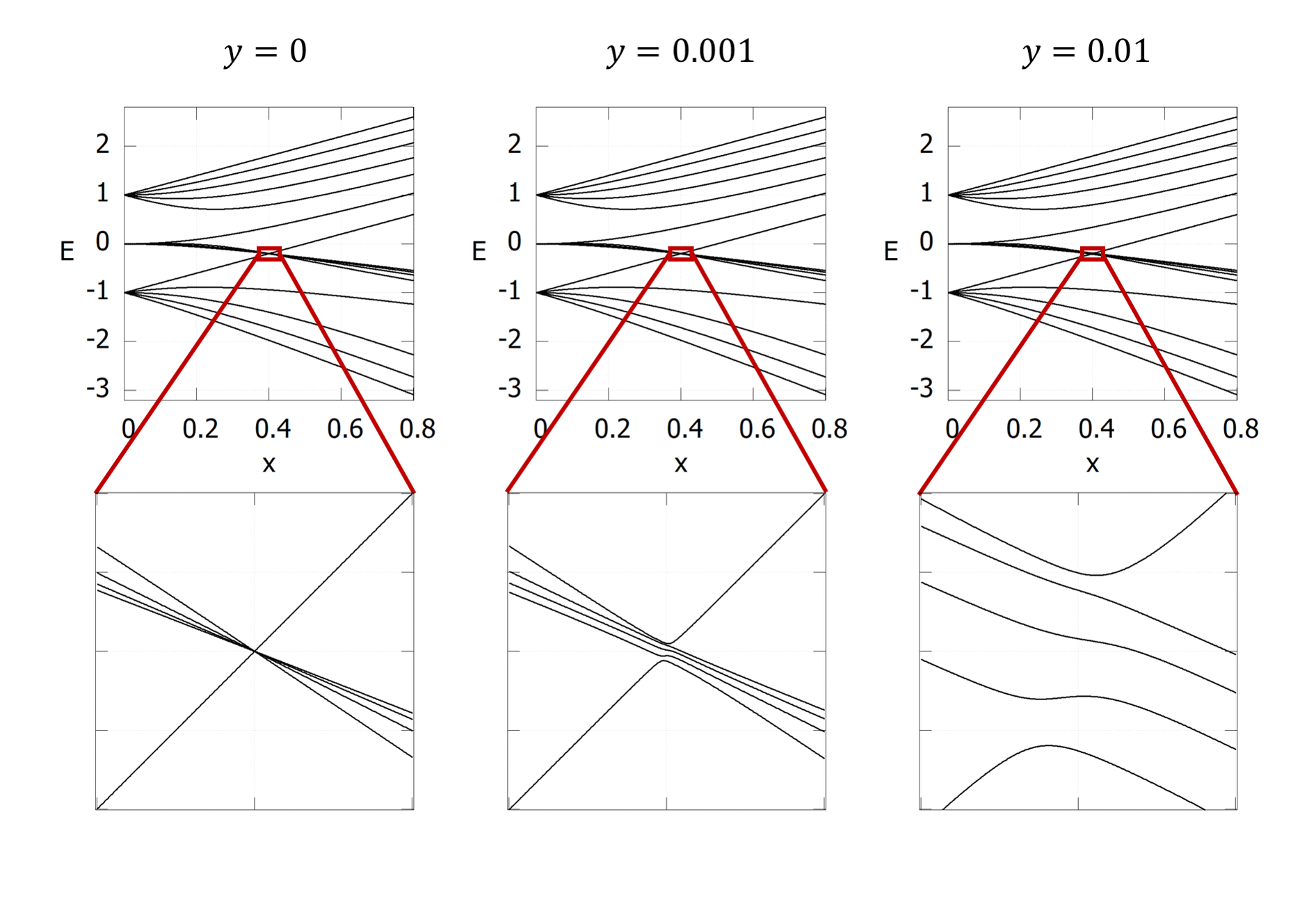}
		\caption{When y=0, the Hamiltonian has a $2L+1$ fold degeneracy. When $y\ne 0$, the crossing turns into anti-crossing.}
		\label{spec2}
	\end{figure}

	The topological properties of Happer model without spin-axis interactions have been discussed in previous sections. In this section, by numerical calculations, we present the effect of spin-axis interactions  on Happer model in Eq. \eqref{Eq:Full-Hamiltonian}, i.e. $y\neq0$,
	\begin{equation}
	H=\hat{\Vec{n}}_{B}\cdot \Vec{S} + x\Vec{S} \cdot \Vec{L} + y \Vec{S} \cdot (3 \hat{\Vec{a}} \hat{\Vec{a}}-\id) \cdot \Vec{S}.
	\end{equation}
	where $\hat{\Vec{a}}$ is a unit vector along the direction of the internuclear axis. The existence of spin-axis interaction removes the degeneracy and converts the level crossings into anti-crossings,  see \figref{spec1} and \figref{spec2}. 	when $y\ne 0$, the component of the total angular momentum $\hat{\Vec{n}}_{B}\!\cdot\!\Vec{J}$ on the direction of magnetic field is no longer conserved for a general $\hat{\Vec{a}}$. The conservation of $\hat{\Vec{n}}_{B}\!\cdot\!\Vec{J}$ only occurs in the case that the direction of magnetic field lies (anti)parallel to the internuclear axis, namely $\hat{\Vec{n}}_{B}=\pm\hat{\Vec{a}}$, which can be easily checked by the commuting relation,
	\begin{align}\label{anti}
	[\hat{\Vec{n}}_{B}\cdot \Vec{J},H(y\ne0)]
	& = 6y\Vec{S}\cdot (\hat{\Vec{n}}_{B}\times \hat{\Vec{a}})\hat{\Vec{a}} \cdot \Vec{S} \begin{cases}
	=0,& \hat{\Vec{n}}_{B} = \pm \hat{\Vec{a}}. \\
	\ne 0,& \hat{\Vec{n}}_{B} \ne \pm \hat{\Vec{a}} .
	\end{cases}
	\end{align}

	
	Without loss of generality, here we mainly focus on  $L=1$. 
	
	In the case of $L=1$ with the spin-axis interaction for $\hat{\Vec{n}}_{B}\neq\pm\hat{\Vec{a}}$, numerical calculations show that the three-fold degeneracy at $x=\frac{2}{3}$ is removed, and the three levels turn to be the anti-crossing, see \figref{spec1}.
	
	Now let us discuss the Chern numbers of all the energy levels.
	
	In our previous calculation for $H(y=0)$, Chern numbers of non-degenerate states are equal to the opposite numbers of $\hat{\Vec{n}}_{B}\cdot \Vec{J}$ on the corresponding states.  However for the anti-crossing case, $\hat{\Vec{n}}_{B}\cdot \Vec{J}$ is no longer conserved, the energy gaps are opened and each energy level hosts different Chern numbers with varying  parameter $x$.
	
\begin{figure}[h!]
		\centering
		\includegraphics[width=0.42\textwidth]{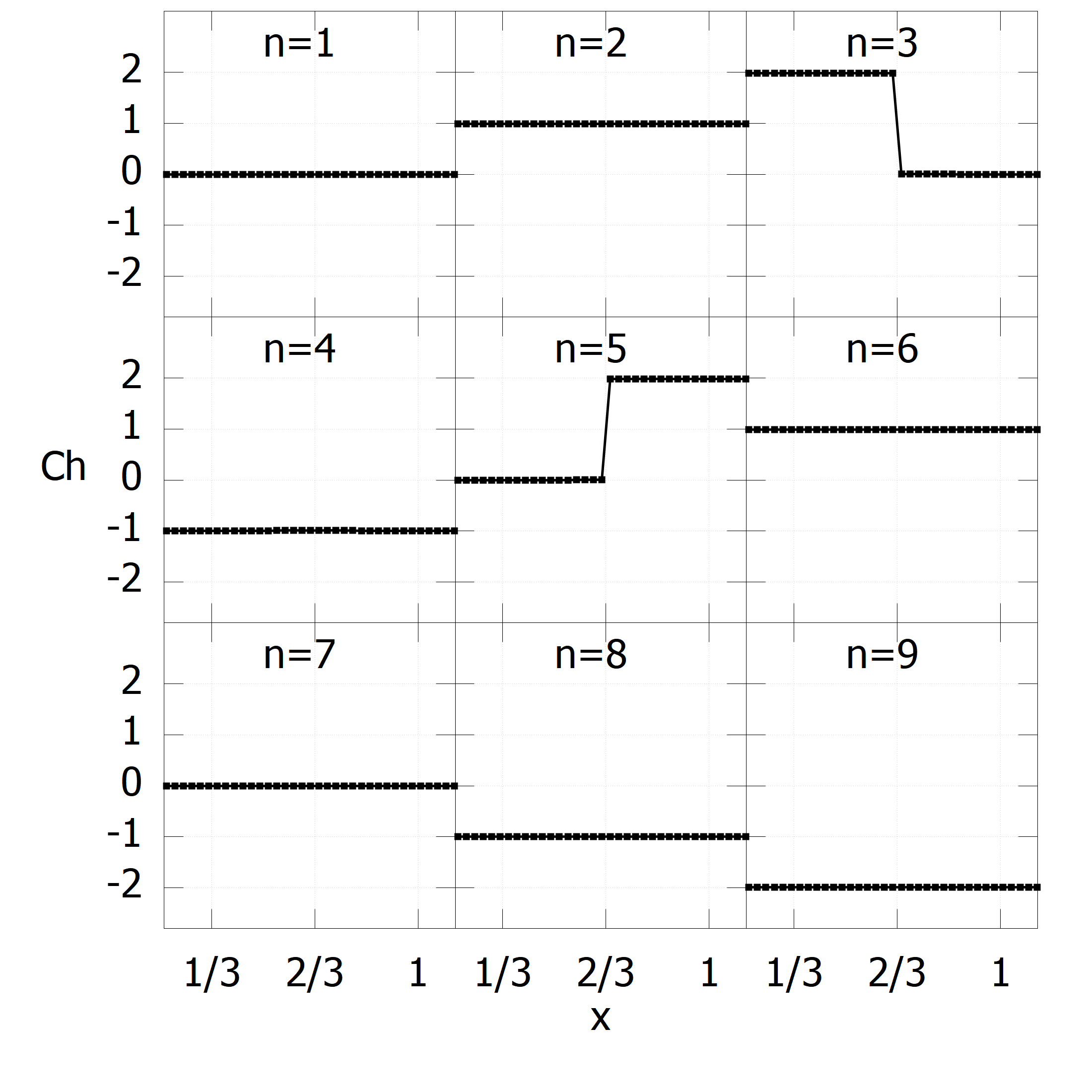}
		\caption{Chern numbers of different energy levels for spin-axis perturbation $y$ in the order of 0.001. The energy level $n=3$ (and $n=5$) host a sudden jump of Chern number near the anti-crossing point $x=\frac{2}{3}$. }
		\label{Ch3}
	\end{figure}

	When the perturbation $y$ is of the order $0.001$, numerical calculations show that the motions of the $\avg{\Vec{S}}$ and $\avg{\Vec{L}}$ are similar to that of $y=0$, only the indexes of energy levels changed. Chern numbers of the states of $n=3,5$  jump at the point of $x=\frac{2}{3}$, see Fig.\ref{Ch3}.
	
\begin{figure}[h!]
		\centering
		\includegraphics[width=0.49\textwidth]{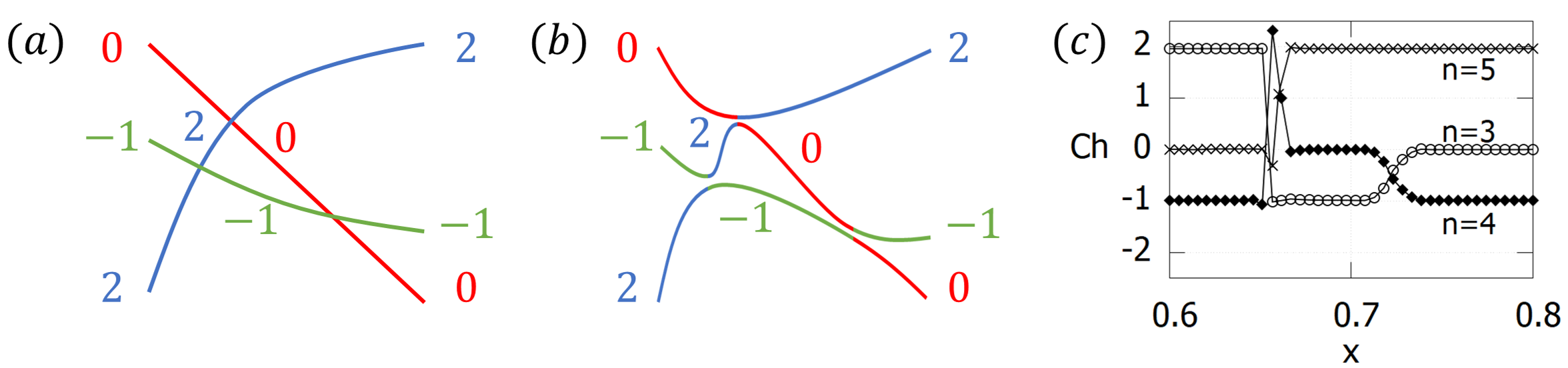}
		\caption{(a) and (b) are schematic plots where the triangle of crossing turns into the anti-crossing with many bumps. The different colors of red, green and blue are related to different Chern numbers of $0$, $-1$ and $2$. (c) are the numerical results of Chern numbers of states of $n=3,4,5$ over $x$. 
		}
		\label{bump1}
	\end{figure}

	When $y$ becomes more prominent, for example near the order of $0.1$, the movements of $\Vec S$ and $\Vec L$ become distorted because of the spin-axis interaction. The degenerate states cross at three different $x$'s, forming a triangle, instead of at the same point like the $y=0.001$ case. This triangle turns into the anti-crossing with many bumps when it meets the condition in Eq.~(\ref{anti}), see Fig.\ref{bump1}.
	
	{Suppose we slowly vary the direction of the axis with nonzero $y$, from linearly dependent to independent to the direction of the magnetic field. From the start, for $y\neq 0$,  the degenerate energy levels (at $y=0$) will intersect at three different points, forming a triangle. Chern number of these energy levels are changed at each crossing. Later, each crossing of the triangle will turn into anti-crossing. Chern number of these energy levels change around these anti crossings, see Fig.\ref{bump1}. }
	
	Near the anti-crossing point  $x=\frac{2}{3}$, the energy gaps are small, hence the Landau-Zener effect\cite{landau1932theory,zener1932non} is evident, namely the state of one energy level can transit to other energy levels with a high probability.  
	
\begin{figure}[h!]
		\centering
		\includegraphics[width=0.44\textwidth]{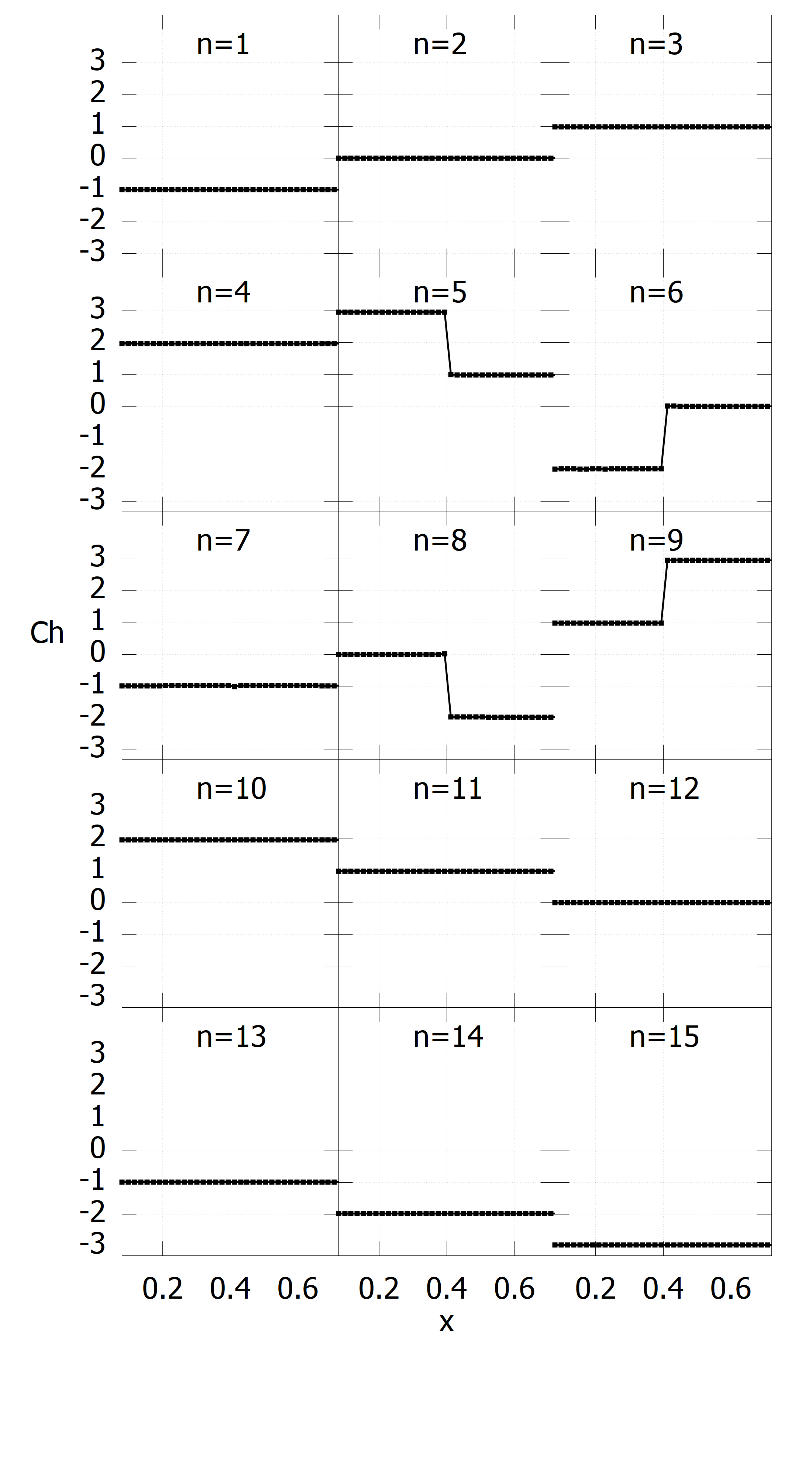}
\caption{Chern numbers of different energy levels for spin-axis perturbation $y$ in the order of 0.001. The energy levels $n=5,6,8,9$ hold sudden jumps of Chern numbers near the anti-crossing point $x=\frac{2}{5}$. }
		\label{Ch4}
	\end{figure}

	Similar analysis can be applied to the $L=2$ case. In Fig.\ref{Ch4}, numerical calcualtions show that the Chern numbers of the states of $n=5,6,8,9$ change near the point $x=\frac{2}{5}$ with the spin-axis perturbation $y$ of the order $0.001$.	The transition of Chern numbers can also be explained by the Landau-Zener transition effect.

	For the dynamical properties of $\Vec{S}$ and $\Vec{L}$, when the perturbation is small, the motions of the $\avg{\Vec{S}}$ and $\avg{\Vec{L}}$ are similar to that of $y=0$, only the indexes of energy levels change. When $y$ becomes more prominent, the movements become distorted.
	
	\section{Comparison with the topological semimetal model}\label{sec:IV}
	In the simplest topological semimetal model, the Hamiltonian near the degenerate point in momentum space takes the following linear formula,
	\begin{equation}
H_{sm}=\vec{\Vec{k}}\cdot \vec{\Vec{\sigma}},
	\end{equation}
where $\vec{\Vec{k}}=(k_x,k_y,k_z)$ represents the momentum vector, and $\vec{\Vec{\sigma}}=(\sigma_x,\sigma_y,\sigma_z)$ are Pauli matrices.   The $k_x=k_y=k_z=0$ point is the so called Weyl point, where the Hamiltonian hosts a double degeneracy. The degeneracy is topological stable and protected by the topological Chern number. Mathematically, the double degeneracy can also be extended to the higher $2n+1$-fold degeneracy: One only needs to change the Pauli matrices for spin-$\frac{1}{2}$ to the higher spin-$n$ operators. For example\cite{hu2018topological}, for $H_{sm}=\vec{\mathbf{k}}\cdot\vec{\mathbf{F}}$, where $\vec{\mathbf{F}}$ is spin-1 operator, the corresponding ``Weyl point'' $k_x=k_y=k_z=0$ should hold triple degeneracy. The topological Chern number is defined via the lowest band state of the system, and can be regarded as the ``magnetic flux'' over a surface. This is nothing but a magnetic monopole phenomena in 3d-momentum space, akin to the Gauss law for the  electrons. Indeed, each band of the Hamiltonian $H_{sm}$ holds a Chern number proportional to the $\vec{\mathbf{k}}$-direction spin quantum number: $-n,-n+1...n-1,n$ for spin-$n$. Summing Chern numbers over all the bands leads to a trivial total Chern number 0.

The Happer model can be compared with the topological semimetal model in the sense of topological numbers\cite{note}. Taking $L=1$ as example. If we substitute the parameters $\Vec{n_B}/x$ in  Eq.~(\ref{Eq:Non-SA-Hamiltonian}) by $\vec{\Vec{k}}=(k_x,k_y,k_z)$, then the transformed Hamiltonian reads $H\longrightarrow H'=\vec{\Vec{k}}\cdot \Vec{S} + \Vec{S} \cdot \Vec{L}$. Projecting the system into the subspace with triple degeneracy,  as obtained in Eq.~(\ref{Hp}), we have
\begin{equation}\label{HPP}
H'_P=\Pi H' \Pi,\quad \Pi=|\psi'_3\rangle\langle\psi'_3|+|\psi'_4\rangle\langle\psi'_4|+|\psi'_5\rangle\langle\psi'_5|,
\end{equation}	
where $|\psi'_{3,4,5}\rangle$ are the three eigenstates corresponding to the three crossing energy levels.

After the projection, we suppose the existence of $H'_p$ in momentum space, then we can make better comparisons between $H'_p$ and the topological semimetal $H_{sm}$ for spin-1.
According to the previous calculation, for $L=1$, the triple degeneracy of $H'_p$ occurs at the configuration $|\vec{\Vec{k}}|=\sqrt{k_x^2+k_y^2+k_z^2}=1/x=\frac{3}{2}$, hence it is more like a ''Weyl sphere'' $\mathbb{S}^2$ for degeneracy, rather than a ``Weyl point'' in usual topological semimetal $H_{sm}$. Fig.~\ref{oppo1} shows that $x=\frac{2}{3}$ is a critical point where the lowest energy band has Chern number 2 for $x<\frac{2}{3}$ and Chern number 0 for  $x>\frac{2}{3}$. In the model $H'_p$,  when $|\vec{\Vec{k}}|=\sqrt{k_x^2+k_y^2+k_z^2}=1/x>\frac{3}{2}$, the lowest energy band Chern number is 2, while for $|\vec{\Vec{k}}|=\sqrt{k_x^2+k_y^2+k_z^2}=1/x<\frac{3}{2}$, the Chern number is 0.  Hence the ``Weyl sphere'' in $H'_p$ acts as the role of critical sphere describing the jump between Chern numbers in the lowest band. 

Another point for comparison is the sum of Chern numbers over all bands. As was present above, for topological semimetal model, the sum of Chern numbers is always 0 for any spin-$n$ operators $\vec{\mathbf{F}}$. Differently, for our model $H'_p$ (L=1) in Eq.~(\ref{HPP}), the sum of Chern numbers is 1. Actually, for any extended $H'_p$ for general $L$, the sum of Chern numbers over $2L+1$ bands is always 1. This can be checked from the exact solution of the Happer model\cite{ge2002Curious}.

	\section{Conclusion and Discussion}
In this paper, we investigate the topological properties of  Happer model under the periodic magnetic field. In the two-atom system, the magnetic field only applies on the triplet spin-dimer $\Vec{S}$. In the simplest case for $y=0$, the Chern numbers for both degenerate and non-degenerate cases are discussed. The Chern number is shown closely connected to the total angular momentum quantum number for each level, rather than the spin-dimer $\Vec{S}$.  Under the perturbation of spin-axis interactions, the level crossing at $x=\frac{2}{2L+1}$ turns out to be the anti-crossing, and the symmetry $\hat{\Vec{n}}_{B}\!\cdot\!\Vec{J}$ is broken. In this case, for each separate energy level, the Chern number varies along the value of $x$. This can be explained by the Landau-Zener effect mainly occuring around the anti-crossing area.

On the other hand, based on the three-fold level crossing at $x=\frac{2}{3}$ in Happer model $H(y=0,L=1)$,  we compare the deformed Hamiltonian $H\longrightarrow H'_p$ in momentum space with the simplest topological spin-1 semimetal model. Akin to the ``Weyl point'' $k_x=k_y=k_z=0$ for triple degeneracy, our model $H'_p$ holds a so called ``Weyl sphere'' $\mathbb{S}^2$ for three-fold degeneracy. More interestingly, the ``Weyl sphere'' $\mathbb{S}^2$ is a critical sphere, where the Chern number jumps, i.e. the phase transition occurs. In electromagnetic language, there is magnetic source on  the  point $k_x=k_y=k_z=0$. For any sphere covering the zero point, when the sphere radius is smaller than $\frac{3}{2}$, the magnetic flux is 0; While for the sphere radius greater than $\frac{3}{2}$, the magnetic flux is proportional to the Chern number 2. This magnetic phenomena has an electronic analogue: Electrostatic shielding. Hence we can call it  ``magnetostatic shielding''--like phenomena in momentum space.

In summary, we show the interesting physical consequences of Happer model with the puzzling degeneracy. Especially, the comparison between the Happer model and the topological semimetal may help in constructing models with more enriching physical consequences.
\bigskip

	\section*{ACKNOWLEDGEMENTS}
	Y. K. Liu and Y. F.  Liu contribute equally to this work.  The authors are grateful to Prof. M.-L. Ge for introducing this research topic and  enlightening discussions. L.-W. Yu is supported by National Natural Science Foundation of China (Grant No. 11905108), China Postdoctoral Science Foundation (Grant No. 2018M641622) and ``the Fundamental Research Funds for the Central Universities'', Nankai University (Grant No. 63191347).	
	

\begin{thebibliography}{20}
\bibitem{note}{Here we only compare the topological numbers of the Happer model and the topological semimetal model. In topological semimetal, the bases are usually $\{a_k^\dag,\, a_k\}$, where $a_k^\dag(a_k)$ represents the fermionic operator in momentum space. However in the Happer model,  we are not able to show the physical interpretation of the basis in momentum space, especially for the projected space spanned by $\{|\psi'_{3}\rangle,\,|\psi'_{4}\rangle,\, |\psi'_{5}\rangle\}$.}
\end{thebibliography}

%

	\newpage
	\onecolumngrid
\begin{center}	{\Large  Supplementary materials}
\end{center}
	\appendix
	\section{Berry phase}
	\label{appdixBerry}
	Consider a smooth and single-valued Hamiltonian $H(\vec{\lambda})$ with the parameter of $\vec{\lambda}$. The parameter $\vec{\lambda}$ evolves periodically with time, namely $\vec{\lambda}(T)=\vec{\lambda}(0)$, so does the state, represented by $\ket{\psi(t)}\bra{\psi(t)}$. But in most cases, the state vector $\ket{\psi(t)}$ will accumulate an extra phase. The state vector at time $T$ will be
	\begin{equation}
	\ket{\psi(T)}=\e^{-\i\alpha_\psi}\ket{\psi(0)}.
	\end{equation}
	
	The phase can be divided into two part: one is dynamic phase, arising from state's time evolution and another is geometrical phase from the variation of the eigenstate with the changing Hamiltonian. In the adiabatic approximation, when states remains an eigenstate of $H(\vec{\lambda})$ at all time with same energy quantum number, the geometrical phase is called Berry phase. Solving the Schr\"{o}dinger equation with $\ket{\psi(t)}=c_n(t)\ket{n,\lambda(t)}$, where $\ket{n,\lambda(t)}$ is the instantaneous energy eigenstate, can give the expression of Berry phase,
	\begin{equation}
	\gamma_n(t)=\int_0^t\i\bra{n,\vec{\lambda}(t)} \frac{\d}{\d t^\prime} \ket{n,\vec{\lambda}(t)}\d t^\prime.
	\end{equation}
	
	Define Berry connection as $\vec{A}^n(\vec{\lambda}) = \i\bra{n,\vec{\lambda}}\nabla\ket{n,\vec{\lambda}}$ and the Berry phase can be written as a integration along a loop in the parameter space,
	\begin{equation}
	\gamma_n(t)=\int_{\vec{\lambda}(0)}^{\vec{\lambda}(t)} \vec{A}^n(\vec{\lambda})\d\vec{\lambda}.
	\end{equation}
	
	The Berry connection $\vec{A}^n$ and Berry phase $\gamma_n$ are gauge variant, while Berry curvature $\vec{F}^n=\nabla\times\vec{A}^n$ is gauge invariant.
	
	\section{Numerical Calculation of Chern Number and Geometrical Phase}
	\label{appdixCh}
	To demonstrate the Numerical Calculation, we will use the non-abelian gauge theory, because the abelian gauge is simpler and can be deduced from non-abelian gauge theory with the reduction of the matrix to number. We will mesh the sphere surface of the parameter space first so that $\theta$ and $\varphi$ will be discrete values in the region of $[0,\pi]$ and $[0,2\pi]$. All the expressions in the non-abelian gauge theory are expressed in the difference or summation form as follows. Berry connection one-form is
	\begin{align}
	A^{kl}(\theta,\varphi) = A_1^{kl}(\theta,\varphi)+A_2^{kl}(\theta,\varphi)= \i\left(\braket{k,\theta,\varphi}{l,\theta+\Delta\theta,\varphi}-\delta_{kl}\right) + \i\left(\braket{k,\theta,\varphi}{l,\theta,\varphi+\Delta\varphi}-\delta_{kl}\right).
	\end{align}
	
	where we apply the orthogonalization and normalization of eigenstate vectors. We divide the matrix of Berry connection one-form into two parts $A_1^{kl}(\theta,\varphi)$ and $A_2^{kl}(\theta,\varphi)$ so that Berry curvature two-form can be expressed as
	\begin{equation}
	F(\theta,\varphi) = A_2(\theta+\Delta\theta,\varphi)-A_2(\theta,\varphi) -A_1(\theta,\varphi+\Delta\varphi)+A_1(\theta,\varphi) + \i\left[A_1(\theta,\varphi),A_2(\theta,\varphi)\right].
	\end{equation}
	
	Chern number and geometrical phase are
	\begin{align}
	\Ch & = \frac{1}{4\pi}\sum_{k;\theta,\varphi\in S^2}F^{kk}(\theta,\varphi),\\
	\gamma(\partial\Sigma) & =\sum_{k;\theta,\varphi\in \Sigma}F^{kk}(\theta,\varphi).
	\end{align}
	
\begin{figure}[h!]
		\centering
		\includegraphics[width=0.8\textwidth]{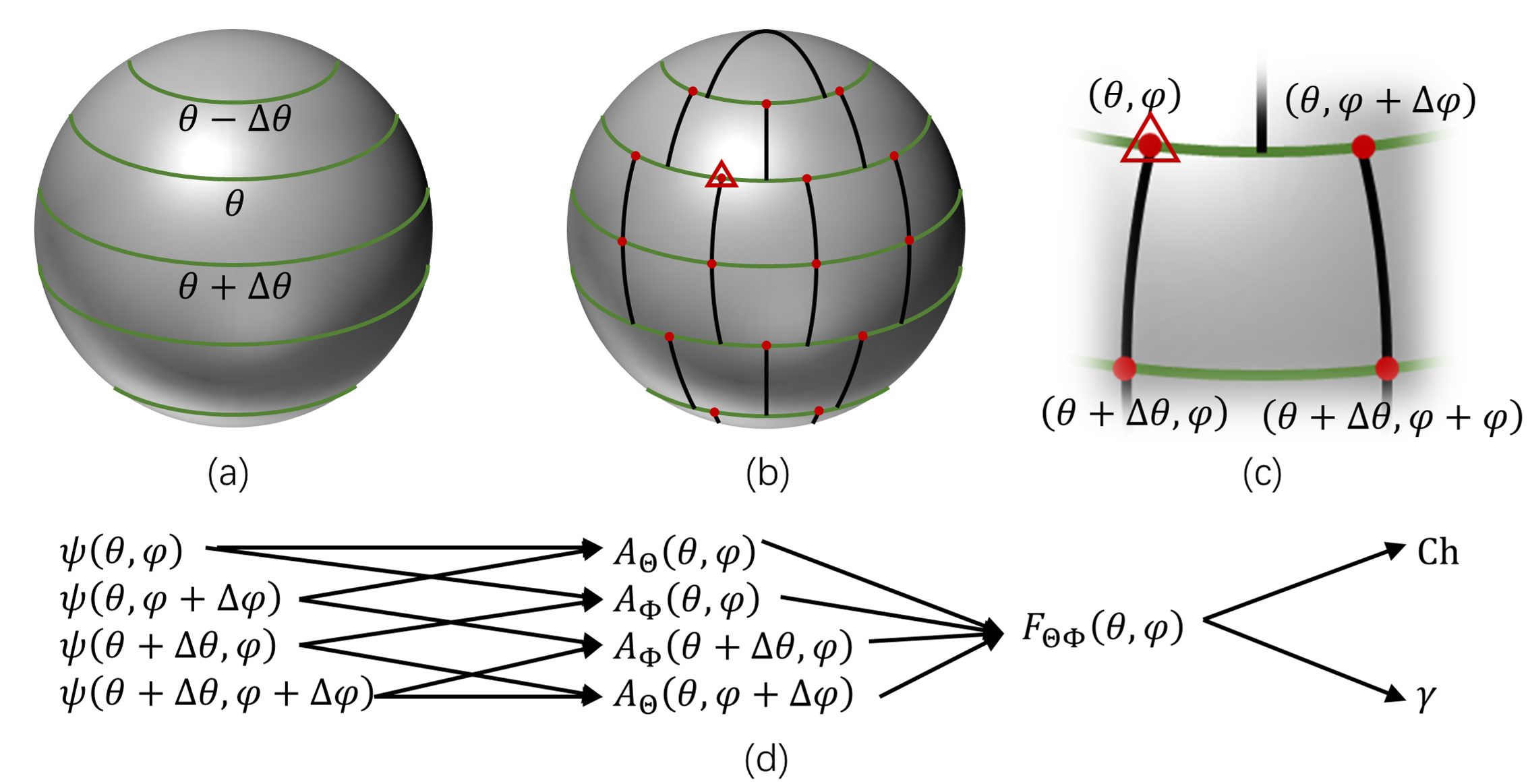}
		\caption{Grid of the parameter space. In (a), we divide the $\theta$ in the parameter space (gray sphere surface) evenly. In (b), we divide the $\varphi$ variably. (c) is a block. The red circles are Representative point and the red circle in a triangle is the Representative point for the block in (c). (d) shows how to get $F_{\Theta\Phi}$ using eigenstate vectors on the 4 corners of the block. To clarify the diagram, the grids in (a) and (b) are so sparse. But in the real calculation, the grid is dense, like dividing each parameter into $~1000$ sections. Since the grid is so dense and average areas of blocks are so small, we can drop the region of $\theta\in [0,\Delta \theta]$, where the representative points are hard to define.}
		\label{grid}
	\end{figure}

	To mesh the sphere surface into average area, $\theta$ is divided evenly and $\varphi$ is divided into different sizes, which is coarse when $\theta\rightarrow 0$ or $\pi$ and fine when $\theta\rightarrow \frac{\pi}{2}$, shown in \ref{grid}. There is a representative point defined as $(\theta,\varphi)$ at the left-up corner for each block. To calculate each Berry curvature $F_{\Theta\Phi}(\theta,\varphi)$, we need the eigenstate vectors on the 4 corners of the block.

	Once we calculate the Berry curvatures of all the discrete points in the parameter space, we can calculate the Chern number and Geometrical phase easily by summation. 
	
	\section{Analytical Calculation of Degenerate States}
	\label{appdixState}
	The calculation detail for degenerate eigenstate vectors for $L=1$ is shown below. The Hamiltonian at the degenerate point $x=\frac{2}{3}$ in the base of $\ket{S_z,L_z}$ is shown as
	
	\begin{equation}
	H(\theta,\varphi)=\left(
	\begin{matrix}
	\cos \theta+\frac{2}{3} & 0 & 0 & \frac{\sin\theta e^{-\i\varphi}}{\sqrt{2}} & 0 & 0 & 0 & 0 & 0 \\
	0 & \cos \theta & 0 & \frac{2}{3} & \frac{\sin\theta e^{-\i\varphi}}{\sqrt{2}} & 0 & 0 & 0 & 0 \\
	0 & 0 & \cos \theta-\frac{2}{3} & 0 & \frac{2}{3} & \frac{\sin\theta e^{-\i\varphi}}{\sqrt{2}} & 0 & 0 & 0 \\
	\frac{\sin\theta e^{\i\varphi}}{\sqrt{2}} & \frac{2}{3} & 0 & 0 & 0 & 0 & \frac{\sin\theta e^{-\i\varphi}}{\sqrt{2}} & 0 & 0 \\
	0 & \frac{\sin\theta e^{\i\varphi}}{\sqrt{2}} & \frac{2}{3} & 0 & 0 & 0 & \frac{2}{3} & \frac{\sin\theta e^{-\i\varphi}}{\sqrt{2}} & 0 \\
	0 & 0 & \frac{\sin\theta e^{\i\varphi}}{\sqrt{2}} & 0 & 0 & 0 & 0 & \frac{2}{3} & \frac{\sin\theta e^{-\i\varphi}}{\sqrt{2}} \\
	0 & 0 & 0 & \frac{\sin\theta e^{\i\varphi}}{\sqrt{2}} & \frac{2}{3} & 0 & -\cos \theta-\frac{2}{3} & 0 & 0 \\
	0 & 0 & 0 & 0 & \frac{\sin\theta e^{\i\varphi}}{\sqrt{2}} & \frac{2}{3} & 0 & -\cos \theta & 0 \\
	0 & 0 & 0 & 0 & 0 & \frac{\sin\theta e^{\i\varphi}}{\sqrt{2}} & 0 & 0 & \frac{2}{3}-\cos \theta \\
	\end{matrix}
	\right).
	\end{equation}
	
	The degenerate energy is $-\frac{1}{3}$, namely 
	\begin{equation}
	\left(H(\theta,\varphi)+\frac{1}{3}\id\right)\psi_\text{deg}=\vec{0}.
	\end{equation}
	
	Now, perform elementary row operations to put the matrix $H(\theta,\varphi)+\frac{1}{3}\id$ into the upper triangular form. The explicit matrix equation is
	
	\begin{equation}
	\left(
	\begin{matrix}
	1 & 0 & 0 & 0 & 0 & 0 & f_{17} & f_{18} & f_{19} \\
	0 & 1 & 0 & 0 & 0 & 0 & f_{27} & f_{28} & f_{29} \\
	0 & 0 & 1 & 0 & 0 & 0 & 0 & f_{38} & f_{39} \\
	0 & 0 & 0 & 1 & 0 & 0 & f_{47} & f_{48} & f_{49} \\
	0 & 0 & 0 & 0 & 1 & 0 & 0 & f_{58} & f_{59} \\
	0 & 0 & 0 & 0 & 0 & 1 & 0 & 0 & f_{69} \\
	0 & 0 & 0 & 0 & 0 & 0 & 0 & 0 & 0 \\
	0 & 0 & 0 & 0 & 0 & 0 & 0 & 0 & 0 \\
	0 & 0 & 0 & 0 & 0 & 0 & 0 & 0 & 0 \\
	\end{matrix}
	\right)\left(
	\begin{matrix}
	\ \\ \ \\ \ \\ \ \\ \psi_\text{deg} \\ \ \\ \ \\ \ \\ \ \\ 
	\end{matrix}
	\right)=\left(
	\begin{matrix}
	0 \\ 0 \\ 0 \\ 0 \\ 0 \\ 0 \\ 0 \\ 0 \\ 0 \\ 
	\end{matrix}
	\right),
	\end{equation}
	
	where $f_{ij}$ represents non-zero elements. The equation can be solved as 
	\begin{equation}
	\psi_\text{deg1}  =\left(
	\begin{matrix}
	f_{19} \\ f_{29} \\ f_{39} \\ f_{49} \\ f_{59} \\ f_{69} \\ 0 \\ 0 \\ -1 \\ 
	\end{matrix}
	\right)
	=\left(\begin{matrix}
	-\frac{2 \sec ^4\left(\frac{\theta }{2}\right)}{9  \e^{\i 4\varphi}}\\
	\frac{2 \sqrt{2} \csc (\theta ) \left(4 \csc ^2(\theta )-4 \cot (\theta ) \csc (\theta )-3\right)}{9  \e^{\i 3\varphi}}\\ 
	\frac{-2 \csc ^2(\theta )+2 \cot (\theta ) \csc (\theta )+3}{3  \e^{\i 2\varphi}}\\
	-\frac{8 \sqrt{2} (\cos (\theta )-1) \csc ^3(\theta )}{9  \e^{\i 3\varphi}}\\ 
	\frac{4 (\cos (\theta )-1) \csc ^2(\theta )}{3  \e^{\i 2 \varphi}}\\ 
	\sqrt{2} \tan \left(\frac{\theta }{2}\right)  \e^{-\i\varphi} \\0\\0\\-1\\
	\end{matrix}\right),
	\end{equation}
	\begin{equation}
	\psi_\text{deg2} =\left(
	\begin{matrix}
	f_{18} \\ f_{28} \\ f_{38} \\ f_{48} \\ f_{58} \\ 0 \\ 0 \\ -1 \\ 0 \\ 
	\end{matrix}
	\right)=\left(
	\begin{matrix}
	\frac{2 \sqrt{2} \csc (\theta ) \left(4 \csc ^2(\theta )-4 \cot (\theta ) \csc (\theta )-3\right)}{9  \e^{\i 3 \varphi}}\\ -\frac{(1-3 \cos (\theta ))^2 \csc ^2(\theta )}{9  \e^{\i 2\varphi}}\\ \frac{2}{3} \sqrt{2} \csc (\theta )  \e^{-\i\varphi} \\ \frac{4 (3 \cos (\theta )-1) \csc ^2(\theta )}{9  \e^{\i 2\varphi}}\\ -\frac{1}{3} \sqrt{2} (3 \cos (\theta )-1) \csc (\theta )  \e^{-\i\varphi} \\ 0\\ 0\\ -1\\ 0\\ 
	\end{matrix}
	\right),
	\end{equation}
	\begin{equation}
	\psi_\text{deg3} =\left(
	\begin{matrix}
	f_{17} \\ f_{27} \\ 0 \\ f_{47} \\ 0 \\ 0 \\ -1 \\ 0 \\ 0 \\ 
	\end{matrix}
	\right)=\left(
	\begin{matrix}
	\frac{-2 \csc ^2(\theta )+2 \cot (\theta ) \csc (\theta )+3}{3  \e^{\i 2\varphi}}\\ \frac{2}{3} \sqrt{2} \csc (\theta )  \e^{-\i\varphi} \\0\\-\frac{1}{3} \sqrt{2} (3 \cos (\theta )+1) \csc (\theta )  \e^{-\i\varphi} \\0\\0\\-1\\0\\0\\
	\end{matrix}
	\right).
	\end{equation}
	
	These are the bases in the 3-dimension degenerate state space and they are smooth in the parameter space. After the orthogonalization and normalization, we can calculate the W-Z phase. 
	
	In the same way, we can calculate the degenerate states of $L=2$ as well. They are 
	\begin{equation}
	\psi_\text{deg1}=\left(\begin{matrix}
	-\frac{1536 \csc ^6(\theta ) \sin ^4\left(\frac{\theta }{2}\right)}{625  \e^{\i 6\varphi}} \\
	-\frac{6 (5 \cos (\theta )-3) \csc ^3\left(\frac{\theta }{2}\right) \sec ^5\left(\frac{\theta }{2}\right)}{625  \e^{\i 5\varphi}} \\
	\frac{8 \sqrt{6} \csc ^2(\theta ) \left(-6 \csc ^2(\theta )+6 \cot (\theta ) \csc (\theta )+5\right)}{125  \e^{\i 4\varphi}} \\
	-\frac{(5 \cos (\theta )+1) \csc \left(\frac{\theta }{2}\right) \sec ^3\left(\frac{\theta }{2}\right)}{25  \e^{\i 3\varphi}} \\
	\frac{-2 \csc ^2(\theta )+2 \cot (\theta ) \csc (\theta )+5}{5  \e^{\i 2\varphi}} \\
	-\frac{384 \sqrt{2} (\cos (\theta )-1) \csc ^5(\theta )}{625  \e^{\i 5\varphi}} \\
	\frac{96 \sqrt{2} (\cos (\theta )-1) \csc ^4(\theta )}{125  \e^{\i 4\varphi}} \\
	-\frac{16 \sqrt{3} (\cos (\theta )-1) \csc ^3(\theta )}{25  \e^{\i 3\varphi}} \\
	\frac{4 \sqrt{2} (\cos (\theta )-1) \csc ^2(\theta )}{5  \e^{\i 2\varphi}} \\
	\sqrt{2}  \e^{-\i\varphi} \tan \left(\frac{\theta }{2}\right) \\
	0 \\
	0 \\
	0 \\
	0 \\
	-1 \\
	\end{matrix}\right),
	\end{equation}
	\begin{equation}
	\psi_\text{deg2}=\left(
	\begin{matrix}
	-\frac{6 (5 \cos (\theta )-3) \csc ^3\left(\frac{\theta }{2}\right) \sec ^5\left(\frac{\theta }{2}\right)}{625  \e^{\i 5\varphi}} \\
	-\frac{24 (3-5 \cos (\theta ))^2 \csc ^4(\theta )}{625  \e^{\i 4\varphi}} \\
	\frac{\sqrt{6} (-40 \cos (\theta )+25 \cos (2 \theta )+31) \csc ^3(\theta )}{125  \e^{\i 3\varphi}} \\
	\frac{-22 \csc ^2(\theta )+10 \cot (\theta ) \csc (\theta )+25}{25  \e^{\i 2\varphi}} \\
	\frac{4}{5} \csc (\theta )  \e^{-\i\varphi} \\
	\frac{96 \sqrt{2} (5 \cos (\theta )-3) \csc ^4(\theta )}{625  \e^{\i 4\varphi}} \\
	-\frac{24 \sqrt{2} (5 \cos (\theta )-3) \csc ^3(\theta )}{125  \e^{\i 3\varphi}} \\
	\frac{4 \sqrt{3} (5 \cos (\theta )-3) \csc ^2(\theta )}{25  \e^{\i 2\varphi}} \\
	-\frac{1}{5} \sqrt{2} (5 \cos (\theta )-3) \csc (\theta )  \e^{-\i\varphi} \\
	0 \\
	0 \\
	0 \\
	0 \\
	-1 \\
	0 \\
	\end{matrix}
	\right),
	\end{equation}
	\begin{equation}
	\psi_\text{deg3}=\left(
	\begin{matrix}
	\frac{8 \sqrt{6} \csc ^2(\theta ) \left(-6 \csc ^2(\theta )+6 \cot (\theta ) \csc (\theta )+5\right)}{125  \e^{\i 4\varphi}} \\
	\frac{\sqrt{6} (-40 \cos (\theta )+25 \cos (2 \theta )+31) \csc ^3(\theta )}{125  \e^{\i 3\varphi}} \\
	-\frac{(1-5 \cos (\theta ))^2 \csc ^2(\theta )}{25  \e^{\i 2\varphi}} \\
	\frac{2}{5} \sqrt{6} \csc (\theta )  \e^{-\i\varphi} \\
	0 \\
	-\frac{16 \sqrt{3} (5 \cos (\theta )-1) \csc ^3(\theta )}{125  \e^{\i 3\varphi}} \\
	\frac{4 \sqrt{3} (5 \cos (\theta )-1) \csc ^2(\theta )}{25  \e^{\i 2\varphi}} \\
	-\frac{1}{5} \sqrt{2} (5 \cos (\theta )-1) \csc (\theta )  \e^{-\i\varphi} \\
	0 \\
	0 \\
	0 \\
	0 \\
	-1 \\
	0 \\
	0 \\
	\end{matrix}
	\right),
	\end{equation}
	\begin{equation}
	\psi_\text{deg4}=\left(
	\begin{matrix}
	-\frac{(5 \cos (\theta )+1) \csc \left(\frac{\theta }{2}\right) \sec ^3\left(\frac{\theta }{2}\right)}{25  \e^{\i 3\varphi}} \\
	\frac{-22 \csc ^2(\theta )+10 \cot (\theta ) \csc (\theta )+25}{25  \e^{\i 2\varphi}} \\
	\frac{2}{5} \sqrt{6} \csc (\theta )  \e^{-\i\varphi} \\
	0 \\
	0 \\
	\frac{4 \sqrt{2} (5 \cos (\theta )+1) \csc ^2(\theta )}{25  \e^{\i 2\varphi}} \\
	-\frac{1}{5} \sqrt{2} (5 \cos (\theta )+1) \csc (\theta )  \e^{-\i\varphi} \\
	0 \\
	0 \\
	0 \\
	0 \\
	-1 \\
	0 \\
	0 \\
	0 \\
	\end{matrix}
	\right),
	\end{equation}
	\begin{equation}
	\psi_\text{deg5}=\left(
	\begin{matrix}
	\frac{-2 \csc ^2(\theta )+2 \cot (\theta ) \csc (\theta )+5}{5  \e^{\i 2\varphi}} \\
	\frac{4}{5} \csc (\theta )  \e^{-\i\varphi} \\
	0 \\
	0 \\
	0 \\
	-\frac{1}{5} \sqrt{2} (5 \cos (\theta )+3) \csc (\theta )  \e^{-\i\varphi} \\
	0 \\
	0 \\
	0 \\
	0 \\
	-1 \\
	0 \\
	0 \\
	0 \\
	0 \\
	\end{matrix}
	\right).
	\end{equation}	
	These are the bases in the 5-dimension degenerate state space and they are smooth in the parameter space. After the orthogonalization and normalization, we can calculate the W-Z phase. 
	
\end{document}